\documentclass[a4paper,11pt]{article}
\pdfoutput=1
\usepackage{jheppub}
\pdfoutput=1

\usepackage{epsf}
\usepackage{epsfig}
\usepackage{subfig}
\usepackage{mathtools}    
\usepackage{float}
\usepackage{multirow}
\usepackage{comment}
\usepackage{nicefrac}
\usepackage{epstopdf}
\usepackage{slashed}
\usepackage{xcolor}
\usepackage{url}
\usepackage{breqn}
\usepackage{amsmath}   
\usepackage{adjustbox} 
\usepackage{chemformula}
\usepackage{multicol}
\usepackage{placeins}
\usepackage{braket}
\usepackage{color}
\usepackage{booktabs}
\usepackage{siunitx} 

\def\nbslash{\rlap{\hspace{0.02cm}/}{\bar n}}

\newcommand{\eps}{\varepsilon}

\graphicspath{ {./Figures/} }
\title{Renormalisation group evolution of the shape function $g_{17}$ in $\bar{B}\to X_s \gamma$ and $\bar{B} \to X_s \ell^+\ell^-$ at subleading power}

\author{Riccardo Bartocci$^{\,a}$, Philipp B\"oer$^{\,b}$ and Tobias Hurth$^{\,a}$} 

\emailAdd{rbartocc@uni-mainz.de}
\emailAdd{philipp.boeer@cern.ch}
\emailAdd{hurth@uni-mainz.de}

\affiliation{${}^a$PRISMA+ Cluster of Excellence \& Institute of  Physics (THEP) \& Mainz Institute for Theoretical Physics, Johannes Gutenberg University, D-55099 Mainz, Germany}
\affiliation{${}^b$CERN, Theoretical Physics Department, CH-1211 Geneva 23, Switzerland}

\abstract{We derive and solve the renormalisation-group (RG) equation of the shape function $g_{17}(\omega,\omega_1;\mu)$, which appears at subleading power in the factorisation of the inclusive decays $\bar{B} \to X_s \gamma$ and $\bar B \to X_s \ell^+\ell^-$. 
Our results provide the first ingredient for a next-to-leading order analysis of the respective resolved-photon ${Q}^{c}_{1} - {Q}_{7\gamma}$ interference contribution, whose current uncertainties are among the largest ones in  both inclusive penguin modes.
As a by-product of our study, we find that the analytic properties of the soft anomalous dimension as well as the jet functions in a factorisation theorem allow for a simplified renormalisation of operators in Heavy-Quark Effective Theory that are composed of fields smeared along distinct light-cone directions.
Their hadronic matrix elements become relevant in various inclusive and exclusive $B$-decays beyond leading power in the heavy-quark and large-energy expansion.
Using these insights, we derive and solve a simpler ``reduced'' RG equation of an amplitude-level soft function that describes the long-distance QCD dynamics for penguin contributions to exclusive $\bar{B}_{d,s} \to \gamma \gamma$ decays, and which has recently been discussed in the literature.
}

\preprint{
\begin{minipage}{4cm}
\small
  \flushright
MITP/24-078 \\
CERN-TH-2024-191 \\
November 25, 2024
\end{minipage}}

\begin{document}

\maketitle

\section{Introduction} 
\label{sec:introduction}
The inclusive penguin modes $\bar B \to X_{s,d} \gamma$ and $\bar B \to X_{s,d} \ell^+\ell^-$ are theoretically very clean modes to indirectly search for new physics (for reviews see~\cite{Hurth:2003vb,Hurth:2010tk,Hurth:2012vp}). 
In particular the semi-leptonic $\bar  B \to X_{s}  \ell^+\ell^-$ modes have the potential to corroborate present tensions~\cite{Huber:2020vup} in corresponding exclusive decays measured by LHCb (see~\cite{Hurth:2023jwr} and references therein), but are affected by independent hadronic uncertainties.
These inclusive penguin modes are golden modes for the Belle II experiment, but measurements of the decay mode $\bar B \to X_{s,d} \ell^+\ell^-$ might be also possible with the LHCb experiment, especially at high-$q^2$~\cite{Amhis:2021oik,Isidori:2023unk,Huber:2024rbw}.
The expected precision of the Belle II measurements~\cite{Belle-II:2018jsg} calls for a systematic reduction of the theoretical errors in the calculations of the decay rates, CP asymmetries and other observables.

Within the local operator product expansion (OPE), the  inclusive penguin modes are dominated by the perturbatively calculable contributions, and subleading contributions start only at the quadratic level in the small expansion parameter $(\Lambda/ m_b)^2$.
However, the OPE  breaks down in these inclusive modes if one considers operators beyond the leading ones.
This breakdown manifests in non-local power corrections, also called resolved contributions, which are characterised by a sub-processes in which the photon couples to light partons instead of connecting directly to the effective weak-interaction vertex~\cite{Lee:2006wn}.

These non-local power corrections can be systematically included using the framework of soft-collinear effective theory (SCET)~\cite{Bauer:2000yr,Bauer:2001yt,Bauer:2002nz,Beneke:2002ph,Beneke:2002ni}.
In case of the inclusive $\bar B \to X_{s} \gamma$ decays, all resolved contributions to $\mathcal{O}(1/m_b)$ have been calculated some time ago at leading order (LO) in perturbation theory~\cite{Benzke:2010js,Benzke:2010tq,Lee:2006wn}.
The analogous contributions to $\bar B \to X_{s ,d} \ell^+ \ell^-$ have also been analysed to $\mathcal{O}(1/m_b)$ more recently~\cite{Hurth:2017xzf,Benzke:2017woq}.
In both cases, an additional uncertainty of $4-5\%$ was found, which represented the largest uncertainty in the prediction of the decay rate of $\bar B \to X_{s} \gamma$~\cite{Misiak:2015xwa} and of the low-$q^2$ observables of $\bar B \to X_{s ,d} \ell^+ \ell^-$~\cite{Huber:2015sra,Huber:2019iqf}.
Numerically, the most important resolved contribution is due to the interference of the current-current four-quark operator $Q^{c}_{1}$ with the electromagnetic dipole operator $Q_{7\gamma}$,
\begin{equation}
    Q_1^q = [\bar{q} \gamma^\mu (1-\gamma_5) b] \, [\bar{s} \gamma_\mu (1-\gamma_5) q] \,, \qquad Q_{7\gamma} = \frac{-e m_b}{8\pi^2} \, \bar{s} \sigma_{\mu\nu} (1+\gamma_5) F^{\mu\nu} b \,,
\end{equation}
with $q = u,c$.
It is worth mentioning that this resolved contribution was first calculated in a local expansion, which implies that the shape function effects were neglected~\cite{Voloshin:1996gw,Ligeti:1997tc,Grant:1997ec,Buchalla:1997ky}. 
It was shown that this leads to an underestimation of that resolved contribution in both penguin inclusive modes~\cite{Benzke:2010js,Benzke:2017woq}.
A new theoretical input~\cite{Gunawardana:2017zix,Gunawardana:2019gep}, namely the derivation of the second moment of the non-perturbative subleading shape function $g_{17}(\omega,\omega_1;\mu)$ in the resolved $Q^{c}_{1} - Q_{7\gamma}$ contribution, reduced the corresponding uncertainty within the $\bar B \to X_s \gamma$ decay rate.
A more recent analysis, however, found a smaller reduction and additional uncertainties~\cite{Benzke:2020htm}: A clear underestimation of the charm-mass dependence and the  missing uncertainty due to the $1/m_b^2$ corrections in~\cite{Gunawardana:2019gep} were shown to be the reasons for this discrepancy~\cite{Benzke:2020htm}.\footnote{As discussed  already in~\cite{Benzke:2020htm}, the difference of the results in~\cite{Gunawardana:2019gep,Benzke:2020htm} are twofold: In~\cite{Benzke:2020htm} the charm-mass dependence was estimated by the running $\overline{\text{MS}}$ mass within the charm penguin diagram, which is naturally calculated at the hard-collinear scale.
Variation of the hard-collinear scale  from $1.3$ GeV to $1.7$ GeV leads to $1.14\, {\rm GeV} \leq m_c \leq 1.26\, {\rm GeV}$. In contrast, the authors in~\cite{Gunawardana:2019gep} took into account in the parametric uncertainties only what leads to the unnaturally small variation of the charm mass, $1.17\, {\rm GeV} \leq m_c \leq 1.23\, {\rm GeV}$. 
Moreover, the authors of~\cite{Gunawardana:2019gep} dropped the $1/m_b^2$ term that is of kinematic origin -- which was included in the initial analysis in~\cite{Benzke:2010js} -- and did also not include an estimate of $1/m_b^2$ contributions in their result. But this $1/m_b^2$ associated with a $1/m_b$ shape function is rather large, but other $1/m_b^2$ corrections (associated with a $1/m_b^2$ shape function) were estimated to be negligible~\cite{Benzke:2010js}. Thus, before the $1/m_b^2$ corrections are not fully estimated~\cite{Benzke2025}, it is most reasonable to use the large $1/m_b^2$ term as conservative estimate of all $1/m_b^2$ corrections in the current result - as it was now done in the initial analysis in~\cite{Benzke:2010js} and in the more recent analysis in ~\cite{Benzke:2020htm}. }
The uncertainties due to the resolved contributions are still among the largest ones in the inclusive penguin modes. 
In total, the resolved contribution due to the interference of $Q^{c}_{1} - Q_{7\gamma}$ is found to be $(5.15 \pm 2.55)\%$ in case of the $\bar B \to X_s \gamma$ mode~\cite{Hurth:2017xzf,Benzke:2017woq}, corresponding to the range $[2.6\%, 7.7\%]$. 
Here, the Voloshin term of $+3\%$ -- which was calculated based on a local expansion~\cite{Voloshin:1996gw,Ligeti:1997tc,Grant:1997ec,Buchalla:1997ky}, and is traditionally subtracted from the resolved contribution -- was added back.
Furthermore, there is a very large scale ambiguity of the order of almost $40\%$, which still has to be taken into account in the result above: In the LO result (including the Voloshin term), the scale of the hard functions (i.e. the Wilson coefficients), is not fixed. 
If one changes their scale from the hard scale to the hard-collinear scale, the final result increases by $40\%$~\cite{Benzke:2020htm}.

This large scale ambiguity and the large charm-mass dependence strongly motivate a systematic calculation of the $\mathcal{O}(\alpha_s)$ radiative corrections within renormalisation-group (RG) improved 
perturbation theory.
Starting point for such an analysis is a factorisation formula for the subleading non-local contributions in terms of hard functions $H$, jet functions $J$ and $\bar{J}$ and soft functions.
At LO this has been established in~\cite{Benzke:2010js}.
More recently, a partial failure of that factorisation formula was healed in~\cite{Hurth:2023paz} by using new refactorisation techniques~\cite{Boer:2018mgl,Liu:2020wbn,Beneke:2022obx,Bell:2022ott}.
The present paper provides the first step towards including the radiative corrections, focusing on the largest resolved contribution from the $Q^{c}_{1} - Q_{7\gamma}$ interference.
The respective LO factorisation formula takes the schematic form
\begin{equation}
   d\Gamma(\bar{B} \to X_s \gamma) \sim H \cdot J \otimes g_{17} \otimes \bar{J}  \,,
\end{equation}
where the $\otimes$ symbol indicates a convolution integral.
We here calculate the one-loop anomalous dimension of the respective soft function in the factorisation theorem -- the subleading shape function $g_{17}(\omega,\omega_1,\mu)$.

\begin{figure}[t]
    \centering
    \includegraphics[width=0.45\textwidth]{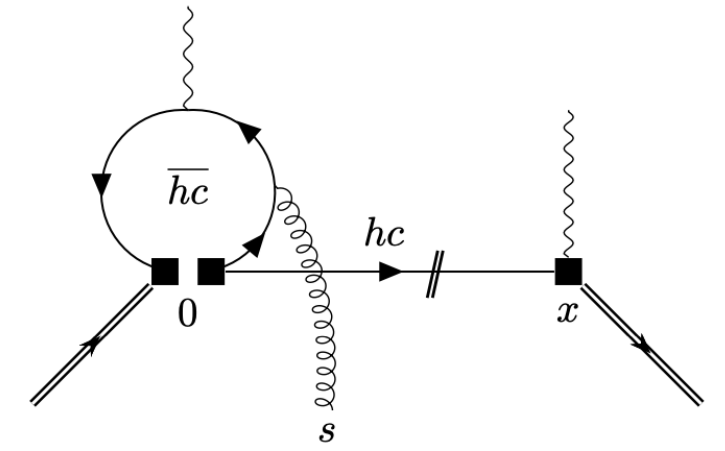}
    \caption{Leading-order contribution to the $\bar{B} \to X_s \gamma$ decay rate from the $Q_1^{c} - Q_{7\gamma}$ interference. Symmetric diagrams are not shown.}
    \label{fig:LO_diagram}
\end{figure}

Conceptually, our analysis also reveals new insights about the renormalisation of a specific set of soft functions. 
Whereas for example the leading-power shape function (and also conventional light-cone distribution amplitudes in exclusive $B$-meson decays),
\begin{align}
   S(\omega;\mu) = \frac{1}{2M_B}
    \int\frac{dt}{2\pi} \, e^{-i\omega t} \, \langle\bar B_v| \big(\bar h_v S_n\big)(tn) \,
    \big(S_n^\dagger h_v\big)(0) |\bar B_v\rangle \,,
\end{align}
whose renormalisation has been studied in~\cite{Bosch:2004th,Bauer:2003pi}, is defined by an operator that only contains soft fields smeared along one common light-cone vector $n^\mu$ in the direction of the energetic $s$-quark, a technical complication in the case of $g_{17}(\omega,\omega_1;\mu)$ is that the underlying operator contains fields smeared along two different light-cones, see the definition below in~\eqref{eq:g17def} and~\eqref{eq:Q17}. 
The reason is that the soft gluon in Fig.~\ref{fig:LO_diagram} couples to the quark loop that converts into the energetic photon in the direction opposite to that of the $s$-quark.

Such soft functions with a dependence on multiple light-cone directions will become relevant in various power-corrections to inclusive and exclusive processes, see e.g.~\cite{Benzke:2010js,Qin:2022rlk,Piscopo:2023opf,Huang:2023jdu,Feldmann:2023plv}, or when QED corrections are included and the external particles are electrically charged~\cite{Beneke:2017vpq,Beneke:2019slt,Beneke:2020vnb,Beneke:2021jhp,Beneke:2021pkl,Beneke:2022msp,Cornella:2022ubo} (see also~\cite{Boer:2023vsg} for a summary of QED corrections in the factorisation approach, and~\cite{Liu:2020eqe,Bodwin:2021cpx,Broggio:2021fnr,Broggio:2023pbu,Beneke:2024cpq,Beneke:2025ufd} for recent analyses of other subleading-power soft functions relevant for the Drell-Yan process as well as $H \to \gamma \gamma$ and $H \to gg$ decays).
In the context of $B$-meson decays, their renormalisation was first rigorously studied in~\cite{Beneke:2022msp} for the QED-generalized $B$-meson light-cone distribution amplitude.
Most importantly, in the present article we find that the anomalous dimension of such a multi-light-cone operator contains terms that are irrelevant in factorisation theorems (and hence for physical quantities), generalizing an observation made in~\cite{Beneke:2022msp}.
This allows one to solve considerably simpler ``reduced'' renormalisation-group (RG) equations.
As an example, we simplify a recent analysis of the RG evolution for an amplitude-level soft function that appears in the factorisation of exclusive $\bar{B}_{d,s}\to\gamma\gamma$ decays, and has recently been studied in~\cite{Huang:2023jdu}.

As a cross-check of this simplification, we have confirmed that all $1/\eps$ singularities cancel at next-to-leading order (NLO) between the hard, (anti-)hard-collinear and soft loops.
Whereas the NLO corrections to the quark jet function and the hard matching coefficients are well-known, we have computed all $1/\eps$ singularities of the anti-hard-collinear two-loop diagrams from gluon attachments to the quark loop.
These corrections turn out to be identical for the inclusive and exclusive process.
We have used the fact that the massless-quark limit can be taken smoothly to simplify the calculation.
This represents a further important step towards a consistent RG analysis, and will be discussed in a forthcoming publication~\cite{Bartocci25}.

We also note that all results on the shape function $g_{17}$ which we derive in the present manuscript can directly be used also in the case of the $\bar B \to X_s \ell^+\ell^-$ mode.

The remainder of the article is organized as follows.
In section 2, we compute the ultraviolet singularities of the soft operator underlying the subleading shape function.
An analytic solution to the corresponding RG equation in momentum-space is presented in section~\ref{subsec:solution}, and some phenomenological implications of the scale evolution are discussed.
In section~\ref{sec:exclusive}, we compare our calculation to a closely related soft function at the amplitude level, which is relevant for penguin contributions to exclusive processes. 
We conclude in section~\ref{sec:conclusion}.

%----------------------------------------------
\section{Renormalisation of the subleading shape function \texorpdfstring{$g_{17}$}{g17}}
The $\bar{B} \to X_s \gamma$ decay rate cannot be expressed as the imaginary part of a $B$-meson forward matrix element of time-ordered operators, because not all possible cuts also contribute to the $b \to s \gamma$ decay.
Instead, it is related to a restricted discontinuity of the forward matrix element of the product of two effective weak Hamiltonians,
\begin{equation}\label{Direct}
   d\Gamma(\bar B\to X_s\gamma)\propto \mbox{Disc}_{\rm \,restr.}\,
   \Big[ i\int d^4 x\,\langle\bar B| {\cal H}_{\rm eff}^\dagger(x)\,
   {\cal H}_{\rm eff}(0) |\bar B\rangle \Big] \,,
\end{equation}
where restricted means that at leading order only the cuts including the photon and the strange quark are considered.
The dominant resolved-photon contribution arises from the interference of the current-current operator $Q_1^{q}$ with the electromagnetic dipole operator $Q_{7\gamma}$.
The leading order contribution to the decay rate is shown in Fig.~\ref{fig:LO_diagram}.

We remind the reader that taking restricted cuts using Cutkosky cutting rules implies that on the left side of the cut one uses standard Feynman rules, whereas on the right side one uses complex-conjugate Feynman rules. 
More precisely, on the left of the cut one uses the standard Feynman rules for propagators $i/(k^2 + i0)$ from time-ordered products as well as for vertices, whereas on the right of the cut one uses complex-conjugate propagators $(-i)/(k^2 - i0)$ from anti-time-ordered products as well as complex-conjugate vertices.
Propagators that cross the cut are set on-shell with a positive-energy constraint,\footnote{See for example eq. (06-128) in the Quantum Field Theory textbook by C. Itzykson and J.-B. Zuber~\cite{itzykson2012quantum}.}
\begin{equation}
\label{eq:cutpropagator}
    \frac{i}{\ell^2 + i0} \to 2 \pi \delta(\ell^2) \theta(\ell_0) \,. 
\end{equation}
The other equivalent option is to consider the absolute-square of the amplitude and sum over the intermediate states $X$ in $
 | \langle \bar B | {\cal H}_{\rm eff} | X \rangle |^2  =
\langle \bar B | {\cal H}_{\rm eff} | X \rangle    \langle \bar B | {\cal H}_{\rm eff} | X \rangle^*$.
A path integral formulation for the evaluation of matrix elements with (restricted) cuts is given by the so-called Keldysh formalism~\cite{Schwinger:1960qe,Keldysh:1964ud} to which we come back below.

The subleading shape-function $g_{17}(\omega,\omega_1;\mu)$, introduced in~\cite{Benzke:2010js}, is the relevant soft function, that captures the non-perturbative low-energy QCD dynamics in the factorisation formula of this resolved contribution, which is part of the restricted discontinuity of the forward matrix element given in~\eqref{Direct}.
The soft function is defined as the Fourier-transformed forward matrix element between two static $\bar{B}$-meson states, 
\begin{align}\label{eq:g17def}
   g_{17}(\omega,\omega_1;\mu) = \frac{1}{2M_B} \int\frac{dr}{2\pi}\,e^{-i\omega_1 r}
    \int\frac{dt}{2\pi}\,e^{-i\omega t} \, \langle\bar B_v| \mathcal{O}_{17}(t,r) |\bar B_v\rangle \,,
\end{align}
of an operator in Heavy-Quark Effective Theory (HQET),
\begin{equation}
\label{eq:Q17}
    \mathcal{O}_{17}(t,r) = \big(\bar h_v S_n\big)(tn)\,
    \nbslash \big(S_n^\dagger S_{\bar n}\big)(0)\,
    i\gamma_\alpha^\perp\bar n_\beta\,
    \big(S_{\bar n}^\dagger\,g_s G_s^{\alpha\beta} S_{\bar n} 
    \big)(r\bar n)\,
    \big(S_{\bar n}^\dagger h_v\big)(0) \,.
\end{equation}
Here, $n^\mu$ is a light-like vector that points into the direction of the energetic $s$-quark (the collinear direction), and, the energetic photon has momentum in the opposite $\bar{n}^\mu$ direction (the anti-collinear direction).
Furthermore, the $S_n$ are soft Wilson lines from the decoupling of the hard-collinear $s$-quark propagator, and correspondingly the $S_{\bar{n}}$ arise from the decoupling of 
anti-hard-collinear propagators in the quark-loop, see Fig.~\ref{fig:LO_diagram} and Fig.~\ref{fig:g17_notation}.
Importantly, the Wilson-lines in both light-cone directions combine to segments of finite length. In the following three subsections we calculate the ultraviolet (UV) singularities of the operator $\mathcal{O}_{17}(t,r)$ using standard HQET Feynman rules from time-ordered products in a $B$-meson forward matrix element without implementing the necessary cuts.
However, once $\alpha_s$ corrections are considered, we have to implement the restricted cuts, because $g_{17}$ is a soft function of a squared amplitude.
We will address this task in Section~\ref{subsec:relevanceofDeltaZ}. 
However, the calculation using time-ordered fields reveals an interesting and surprising feature: the anomalous dimension contains terms that are irrelevant in the factorisation theorem.
We use this observation to simplify an amplitude-level soft function for an exclusive decay in Section~\ref{sec:exclusive}.

\subsection{UV singularities of \texorpdfstring{$\mathcal{O}_{17}$}{O17}}
\label{sec:renormalization}
We compute the ultraviolet (UV) singularities of the momentum-space operator
\begin{equation}
    \widetilde{\mathcal{O}}^{\rm (bare)}_{17}(\omega,\omega_1) = \int\frac{dr}{2\pi}\,e^{-i\omega_1 r}
    \int\frac{dt}{2\pi}\,e^{-i\omega t} \, \mathcal{O}^{\rm (bare)}_{17}(t,r)
\end{equation}
at one-loop order, and extract its $Z$-factor in the $\overline{\text{MS}}$-scheme from
\begin{equation}
  \widetilde{\mathcal{O}}^{\rm (bare)}_{17}(\omega,\omega_1) = \int \! d\omega^{\prime} \int \! d\omega_1^{\prime} \, Z_{17}^{-1}(\omega,\omega_1,\omega',\omega'_1;\mu) \,  \widetilde{\mathcal{O}}^{\rm (ren)}_{17}(\omega,\omega_1;\mu) \,.
\end{equation}
Despite the non-perturbative nature of the shape function $g_{17}$, the UV-singularities of the defining operator can safely be computed in perturbation theory using partonic external states. 
More specifically, in the following we compute one-loop corrections to the matrix element $\langle \widetilde{\mathcal{O}}^{\rm (bare)}_{17}(\omega,\omega') \rangle \equiv \langle h_v(k,s) g(k_g,\lambda)| \widetilde{\mathcal{O}}^{\rm (bare)}_{17}(\omega,\omega') | h_v(k',s') \rangle$.
As a consequence of the space-time arguments of the fields in~\eqref{eq:Q17}, it suffices at tree-level to set the residual heavy-quark momentum to $k^\mu = \frac12 (nk) \bar{n}^\mu \equiv \frac12 k_+ \bar{n}^\mu$, and similarly the gluon momentum to $k_g^\mu = \frac12 (\bar{n} k_g) n^\mu \equiv \frac12 k_{g-} n^\mu$.
At $\mathcal{O}(\alpha_s)$, however, we keep the parton momenta slightly off-shell whenever necessary to regularize potential infrared (IR) divergences in the loop integrals.
Further, we perform the calculation with a ``physical gluon'', i.e. its polarization vector $\epsilon^a_\mu(k_g,\lambda)^*$ is chosen to be perpendicular to $n^\mu$ and $\bar{n}^\mu$. 
\begin{figure}[t]
    \centering
     \includegraphics[width=0.8\textwidth]{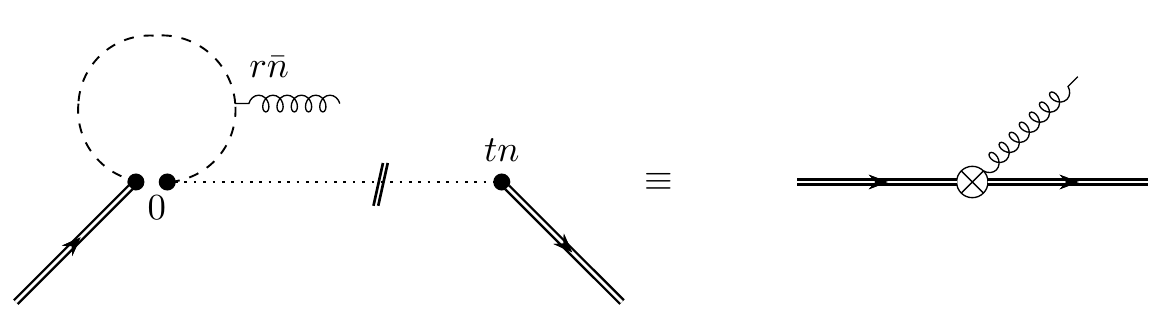}
    \caption{The left-hand side shows the non-local space-time structure of the soft fields and Wilson lines as they appear in the effective operator $\mathcal{O}_{17}$ in~\eqref{eq:Q17}. Here, dashed lines denote soft Wilson lines in the $\bar{n}^\mu$ direction from the decoupling of anti-hard-collinear quark fields, and similarly the dotted line represents a soft Wilson line in the $n^\mu$ direction. Throughout this paper, we simply represent the non-local field configuration through an effective vertex denoted by the $\otimes$ symbol on the right-hand side of the equation. Note that at higher orders, more than one gluon can be emitted from the effective vertex due to the presence of the Wilson lines.}
    \label{fig:g17_notation}
\end{figure}

At tree-level, the partonic matrix element follows from the Feynman rule
\begin{equation}
\label{eq:FR1}
  \vcenter{\hbox{\vspace{+0.2cm} \includegraphics[width=0.22\textwidth]{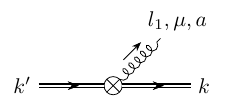}}} = \, \delta(\omega-k_+)\delta(\omega_1-\ell_{1-}) g_s t^a \nbslash (\bar{n}^\mu\slashed{\ell}_{1\perp}-\gamma_\perp^\mu \ell_{1-}) \,,
\end{equation}
and evaluates to
\begin{equation}
\label{eq:LOME}
    \langle \widetilde{\mathcal{O}}^{\rm (bare)}_{17}(\omega,\omega_1) \rangle^{(0)} = - \omega_1 \delta(\omega_1 - k_{g-}) \delta(\omega - k_+) \, [\bar{h}_v \nbslash \slashed{A}_\perp h_v] \,,
\end{equation}
where now the objects in the square brackets represent on-shell spinors for the heavy quarks, and a polarization vector for the gluon, $A_\mu = g_s \epsilon^a_\mu(k_g,\lambda)^* t^a$.
As depicted in Fig.~\ref{fig:g17_notation}, the $\otimes$ symbol used in~\eqref{eq:FR1} and in the following represents the non-local operator $\mathcal{O}_{17}$, where the non-localities along the different light-cones lead to the $\delta$-functions in the momentum-space Feynman rule.
Due to the spin-symmetry of soft-gluon interactions with heavy-quark fields, the spinor product remains unchanged at the loop level. 
The prefactor $(-\omega_1)$ arises from the derivatives in the gluon field-strength tensor.

\begin{figure}[t]
    \centering
     \includegraphics[width=1\textwidth]{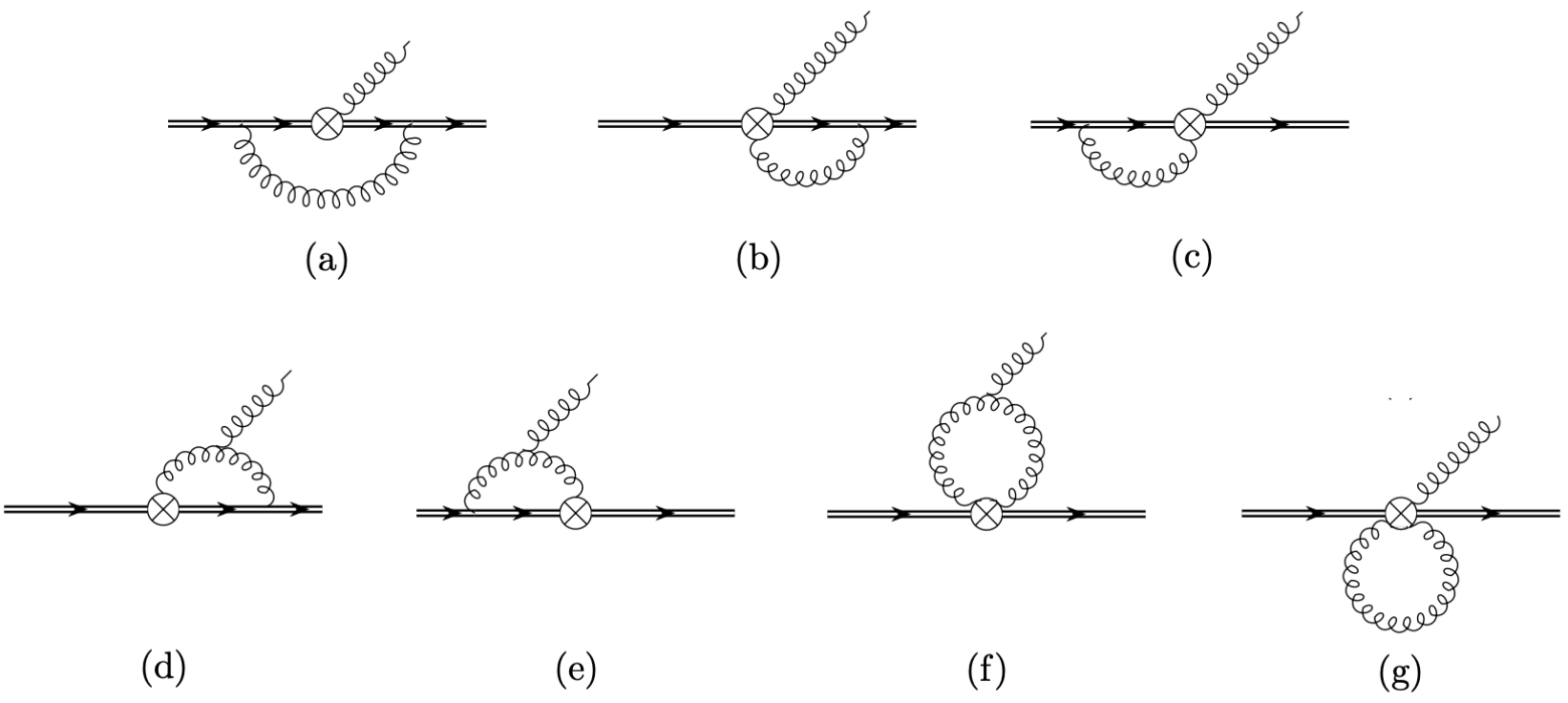}
    \caption{One-loop diagrams that contribute to the partonic matrix element $\langle \widetilde{\mathcal{O}}_{17} \rangle$. The square denotes the insertion of the operator. Self-energy contributions are not shown.}
    \label{fig:g17_diagrams}
\end{figure}
All one-loop graphs that contribute to the UV singularities in Feynman gauge are shown in Fig.~\ref{fig:g17_diagrams}.
We note that the external gluon does not couple directly to the heavy-quark line due to its perpendicular polarisation.
Diagrams without the additional gluon in the final state vanish due to rotational symmetry, i.e. $\mathcal{O}_{17}$ does not mix into operators with only two quark fields.
Additionally, the gluon exchange between the two heavy-quark lines in diagram $(a)$ is UV-finite.
Evaluating the diagrams in Feynman gauge demands Feynman rules for up to three gluon emissions from the operator $\mathcal{O}_{17}$. 
They arise from the gluon field-strength tensor as well as from the various Wilson lines in~\eqref{eq:Q17}.
For the two-gluon emission Feynman rule one finds
\begin{align}
\label{eq:FR2}
  &\, \vcenter{\hbox{\vspace{+0.2cm} \includegraphics[width=0.22\textwidth]{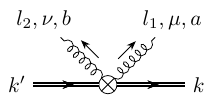}}} = \, -\delta(\omega - k_+) \delta(\omega_1 - \ell_{1-}-\ell_{2-}) \, g_s^2 i f^{abc} t^c \nbslash \gamma_\perp^\mu \bar{n}^\nu \\
    &\qquad + \delta(\omega_1 - \ell_{2-}) \, \frac{n^\mu}{\ell_{1+}} \Big( \delta(\omega - k_+ - \ell_{1+}) - \delta(\omega - k_+)\Big) \, g_s^2 t^a t^b \nbslash (\bar{n}^\nu \slashed{\ell}_{2\perp} - \gamma_\perp^\nu \ell_{2-}) \nonumber \\[0.2em]
    &\qquad - \delta(\omega - k_+) \, \frac{\bar{n}^\mu}{\ell_{1-}} \, \Big( \delta(\omega_1 - \ell_{1-} - \ell_{2-}) - \delta(\omega_1 - \ell_{2-}) \Big) \, g_s^2 i f^{abc} t^c \nbslash 
 (\bar{n}^\nu \slashed{\ell}_{2\perp} -  \gamma_\perp^\nu \ell_{2-}) \nonumber \\[0.2em]
    &\qquad + (l_1, \mu,a) \leftrightarrow (l_2, \nu,b) \nonumber \,.
\end{align}
Here the first line describes two gluons that are emitted from the field-strength tensor $G_s^{\alpha\beta}$, whereas the second and third line describe one gluon from $G_s^{\alpha\beta}$ and one from the finite-length Wilson lines in the $n^\mu$ and $\bar{n}^\mu$ direction, respectively.

For diagram $(g)$ we also need the three-gluon emission Feynman rule, which is a quite lengthy expression.
However, most of the contractions vanish in Feynman gauge, or due to rotational symmetry, and here we only quote the relevant piece that gives a non-vanishing contraction in diagram (g):
\begin{align}
\label{eq:FR3}
  &\, \vcenter{\hbox{\vspace{+0.2cm} \includegraphics[width=0.22\textwidth]{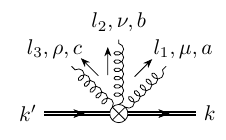}}} \ni \, -\frac{n^\mu}{\ell_{1+}}\Big(\delta(\omega-k_+-\ell_{1+})-\delta(\omega-k_+)\Big)ig_s^3f^{bcd}t^at^d \nbslash \\
&\times \Big[\delta(\omega_1-\ell_{2-}-\ell_{3-})\gamma_\perp^\nu\bar{n}^\rho + \frac{\bar{n}^\nu}{\ell_{2-}}\Big(\delta(\omega_1-l_{2-}-\ell_{3-})-\delta(\omega_1-l_{3-})\Big) \Big(\slashed{\ell}_{3\perp}\bar{n}^\rho-\gamma_\perp^\rho \ell_{3-}\Big)\Big]\nonumber\\
  &+ \text{permutations} \nonumber \,.
\end{align}

We regulate UV-divergences in dimensional regularization, i.e. by evaluating the loop integrals in $d = 4 - 2\eps$ space-time dimensions. 
After the integrations have been carried out, the respective expressions need to be expanded around $\eps = 0$ in the distribution sense.
Besides the standard plus-distributions,
\begin{equation}
    \int \! d\omega' \, \big [ \dots]_+ \, f(\omega') \equiv \int \! d\omega' \, \big [ \dots] \, (f(\omega')-f(\omega)) \,,
\end{equation}
one also needs the modified plus-distributions,
\begin{equation}
    \int \! d\omega' \, \big [ \dots]_{\oplus/\ominus} \, f(\omega') \equiv \int \! d\omega' \, \big [ \dots] \, (f(\omega')-\theta(\pm \omega')f(\omega)) \,,
\end{equation}
which arise because the variables $\omega$ and $\omega_1$ can take both positive and negative values.
It turns out to be convenient to define
\begin{eqnarray}
\label{eq:FandGdistributions}
F^{>}(\omega, \omega^{\prime}) & = &
\left [ \frac{\omega \, \theta (\omega^{\prime} - \omega)} {\omega^{\prime} \,  (\omega^{\prime} - \omega)}   \right ]_{+}
+ \left [ \frac{\theta (\omega - \omega^{\prime})}  {\omega - \omega^{\prime}}  \right ]_{\oplus} \,,
\nonumber \\
F^{<}(\omega, \omega^{\prime})  & = &
\left [ \frac{\omega \, \theta (\omega - \omega^{\prime})} {\omega^{\prime} \,  (\omega- \omega^{\prime})}   \right ]_{+}
+ \left [ \frac{\theta (\omega^{\prime} - \omega)}  {\omega^{\prime} - \omega}  \right ]_{\ominus}  \,,
\nonumber \\
G^{>}(\omega, \omega^{\prime})  & = & (\omega + \omega^{\prime}) \,
\left [ \frac{\theta (\omega^{\prime} - \omega)} {\omega^{\prime} \,  (\omega^{\prime} - \omega)}   \right ]_{+}
- i  \pi  \delta(\omega - \omega^{\prime}) \,,
\nonumber \\
G^{<}(\omega, \omega^{\prime})  & = &   (\omega + \omega^{\prime}) \,
\left [ \frac{ \theta (\omega - \omega^{\prime})} {\omega^{\prime} \,  (\omega- \omega^{\prime})}   \right ]_{+}
+ i  \pi \delta(\omega - \omega^{\prime})  \,,
\end{eqnarray}
as well as the linear combinations
\begin{equation}
\label{eq:H_distributions}
H_{\pm}(\omega, \omega^{\prime}) =  \theta( \pm \omega)  F^{> (<)} (\omega, \omega^{\prime})
+  \theta( \mp \omega)  G^{< (>)} (\omega, \omega^{\prime}) \,,
\end{equation}
which have been introduced in the study of QED corrections to the $B$-meson light-cone distribution amplitude in~\cite{Beneke:2022msp}.

%----------------------------------------------
\subsection{Abelian contributions}
\label{subsec:CFterms}
In the Abelian limit, all soft Wilson lines in the $\bar{n}$-direction in~\eqref{eq:Q17} cancel.
From the spin-symmetry of HQET, it follows that the UV singularities of $\mathcal{O}_{17}$ proportional to the color factor $C_F$ coincide with those of the leading shape function, computed e.g. in~\cite{Bosch:2004th}.
The contribution to the $Z$-factor, which arises from diagrams $(b)$ and $(c)$ in Fig.~\ref{fig:g17_diagrams} (recall that diagram $(a)$ is UV-finite in Feynman gauge), reads
\begin{align}
\label{eq:CFZfactor}
    Z&(\omega,\omega_1,\omega',\omega'_1;\mu)  = \delta(\omega_1 - \omega'_1) \delta(\omega-\omega') \\[0.2em]
    &+ \frac{\alpha_s C_F}{4\pi} \delta(\omega_1 - \omega'_1) \bigg\{ \left(\frac{2}{\eps^2} + \frac{4}{\eps} \ln \frac{\mu}{\bar{\Lambda}-\omega} - \frac{2}{\eps} \right) \delta(\omega-\omega')
    - \frac{4}{\eps} \left[ 
\frac{\theta(\omega'-\omega)}{\omega'-\omega} \right]_+^{(\bar{\Lambda})} \bigg\}
+ \mathcal{O}(\alpha_s^2) \,, \nonumber
\end{align}
where $\bar{\Lambda} = M_B - m_b$ is the difference of the $B$-meson mass and the $b$-quark pole mass.
The superscript $(\bar{\Lambda})$ on the plus-distribution indicates that the integral in $\omega'$ is restricted to the interval $(-\infty, \bar{\Lambda}]$,
\begin{equation}
    \int_{-\infty}^{\bar{\Lambda}} \! d\omega' \, \big [ \dots]_+ \, f(\omega') \equiv \int_{-\infty}^{\bar{\Lambda}} \! d\omega' \, \big [ \dots] \, (f(\omega')-f(\omega)) \,,
\end{equation}
which is also the reason for the logarithmic dependence on $\bar{\Lambda}$.

In the following, we prefer to use an alternative representation that is independent of the IR parameter $\bar{\Lambda}$, but is equivalent to~\eqref{eq:CFZfactor} when acting on a function supported on aforementioned interval. 
This expression is derived by computing diagrams $(b)$ and $(c)$ by means of the residue theorem.
We choose $\ell$ to be the gluon momentum that flows into the square vertex.
Picking up the residues in the $\ell_-$-component and performing the integration over the perpendicular directions, one obtains the following integral for the contribution to the $Z$-factor from diagram $(b)$:
\begin{align}
\label{eq:diagrambCF}
    &-\frac{\alpha_s C_F e^{\eps \gamma_E} \mu^{2\eps}}{2 \pi} \delta(\omega_1 -\omega'_1) \Gamma(\eps) \int_0^\infty \! d\ell_+ \, \ell_+^{-2\eps} \, \frac{1}{\ell_+ + \Delta}\Big(\delta(\omega - \omega' + \ell_+) - \delta(\omega - \omega')\Big) \nonumber \\[0.2em]
    = \, &-\frac{\alpha_s C_F}{2 \pi} \delta(\omega_1 -\omega'_1) \frac{1}{\eps} \bigg\{\frac{\theta(\omega'-\omega)}{\omega'-\omega + \Delta} - \delta(\omega - \omega') \left(\frac{1}{2\eps} - \ln \frac{\Delta}{\mu} \right) \bigg\} + \mathcal{O}(\eps^0)\,.
\end{align}
Here $\Delta$, with $\text{Im} \,\Delta >0$, is a shift in the Wilson-line propagator that arises from off-shell lines in QCD diagrams, and regularizes IR divergences from $\ell_+ \to 0$ in the two individual terms. 
We note that, because the Wilson lines are of finite length, the UV-poles in $\eps$ in~\eqref{eq:diagrambCF} are independent of $\Delta$ in the limit $\Delta \to 0$ once consistently expanded in the distribution sense.\footnote{We note that in some cases the Wilson lines do not combine to finite segments, but instead extend to infinity. The UV poles of the operators then depend logarithmically on such a regulator, hence spoiling its renormalisation. In addition, the regulator-dependent contributions are also gauge-dependent~\cite{Beneke:2024cpq}. An additional rearrangement needs to be performed to remove these contributions from ``charges at infinity'', see e.g.~\cite{Beneke:2019slt,Beneke:2020vnb,Beneke:2021jhp,Beneke:2022msp,Beneke:2024cpq}, thereby restoring gauge-invariance and renormalizability.}
The expansion of the second line of~\eqref{eq:diagrambCF} in terms of (modified) plus-distributions has been discussed in detail in~\cite{Beneke:2022msp}.
The result reads
\begin{align}
\label{eq:diagrambCFdist}
    \frac{\alpha_s C_F}{2 \pi} \delta(\omega_1 -\omega'_1) \bigg\{ &\delta(\omega-\omega') \left( \frac{1}{2\eps^2} + \frac{1}{2\eps}\ln \frac{\mu^2}{\omega^2} \right) \nonumber \\[0.2em]
    - \, &\frac{1}{\eps}\theta(\omega) \left[\frac{\theta(\omega'-\omega)}{\omega'(\omega'-\omega)}\right]_+ \omega' 
    - \frac{1}{\eps}\theta(-\omega) \left[\frac{\theta(\omega'-\omega)}{\omega'-\omega}\right]_\ominus 
     \bigg\} \,,
\end{align}
and is indeed independent of $\Delta$.
Diagram $(c)$ yields the same result, and, after taking into account the divergent one-loop contribution from the heavy-quark wave-function renormalisation constant,\footnote{Note that this constant contributes to $Z^{(1)}_{17}$ with a negative sign due to $Z_h^{1/2} h_v^{\rm ren} = h_v^{\rm bare}$.}
\begin{equation}
    Z_h^{(1)} -1 = \frac{\alpha_s C_F}{2\pi\eps},
\end{equation}
we obtain for the Abelian part of the $Z$-factor
\begin{align}
\label{eq:CFZfactorfinalnocutoff}
    Z_{17}^{(1)}(\omega,\omega_1,\omega',\omega'_1;\mu)\Big\vert_{C_F} = \, &\frac{\alpha_s C_F}{4\pi} \delta(\omega_1 - \omega'_1) \bigg\{ \left( \frac{2}{\eps^2} + \frac{2}{\eps}\ln \frac{\mu^2}{\omega^2} - \frac{2}{\eps}\right) \delta(\omega-\omega')  \\[0.2em]
    &- \frac{4}{\eps}\theta(\omega) \left[\frac{\theta(\omega'-\omega)}{\omega'(\omega'-\omega)}\right]_+ \omega' 
    - \frac{4}{\eps}\theta(-\omega) \left[\frac{\theta(\omega'-\omega)}{\omega'-\omega}\right]_\ominus \bigg\} \,. \nonumber
\end{align}
Here the superscript $(1)$ denotes that we work at $\mathcal{O}(\alpha_s)$. 
It is straightforward to see that this form is equivalent to~\eqref{eq:CFZfactor} by employing the following identities
\begin{align}
    \theta(\omega) \left[\frac{\theta(\omega'-\omega)}{\omega'(\omega'-\omega)}\right]_+ \omega' &= \theta(\omega)   \left[\frac{\theta(\omega'-\omega)}{\omega'-\omega}\right]_+^{(\bar{\Lambda})} 
    +
    \theta(\omega) \delta(\omega-\omega') \ln \frac{\bar{\Lambda}-\omega}{\omega} \,, \nonumber \\[0.2em]
    \theta(-\omega) \left[\frac{\theta(\omega'-\omega)}{\omega'-\omega}\right]_\ominus &= \theta(-\omega)   \left[\frac{\theta(\omega'-\omega)}{\omega'-\omega}\right]_+^{(\bar{\Lambda})} 
    + \theta(-\omega) \delta(\omega-\omega') \ln \frac{\bar{\Lambda} - \omega}{-\omega} \,,
\end{align}
which hold if the test functions have support on the interval $\{\omega,\omega'\} \in (-\infty, \bar{\Lambda}]$.
For the two expressions in~\eqref{eq:CFZfactor} and~\eqref{eq:CFZfactorfinalnocutoff} to be equivalent, it is crucial that only $\theta$-functions that enforce $\omega'>\omega$ appear inside the plus distributions.
This guarantees that a function with support $\bar{\Lambda} > \omega'$ is mapped onto a function with support $\bar{\Lambda} > \omega$.

%----------------------------------------------
\subsection{Non-Abelian contributions}
\label{subsec:CAterms}
Next, we examine the non-Abelian contributions, which arise from all diagrams shown in Fig.~\ref{fig:g17_diagrams} except $(a)$. 
For transparency, we separately discuss the contributions that arise from gluons emitted from the field-strength tensor, from the Wilson lines in the $n^\mu$ direction or the $\bar{n}^\mu$ direction. 
The contraction of Wilson lines in the two light-cone directions in diagram $(g)$ is discussed on its own. 

Setting all Wilson lines to unity leaves diagrams $(d)$ and $(e)$, as well as $(b)$, $(c)$, and $(f)$, but using only the first line of the Feynman rule in~\eqref{eq:FR2} for the latter three. 
In this case, the UV poles of diagrams $(d)$ and $(b)$ cancel.
Similarly, the contribution from diagram $(c)$ is canceled by a piece from diagram $(e)$, but a UV-divergent contribution from $(e)$ remains.
This piece can simply be expanded in the dimensional regulator $\eps$ without the need of introducing plus-type distributions.
The contribution to the $Z$-factor reads
\begin{equation}
\label{eq:Zfactorfirstpiece}
    \frac{\alpha_s C_A}{8\pi \eps} \delta(\omega-\omega') \frac{\omega_1}{(\omega'_1)^2} \big[ \theta(\omega_1)\theta(\omega'_1 - \omega_1) - \theta(-\omega_1) \theta(\omega_1 - \omega'_1) \big] \,.
\end{equation}
Lastly, the contribution from diagram $(f)$ is purely local.
Taking into account the symmetry factor $2$, one finds
\begin{equation}
    \frac{3 \alpha_s C_A}{8\pi\eps} \delta(\omega-\omega') \delta(\omega_1 - \omega'_1) \,.
\end{equation}

Next, we investigate diagrams $(b)$, $(c)$ and $(f)$ using the second line of~\eqref{eq:FR2}, as well as diagram $(g)$ using only the first term in the second line of~\eqref{eq:FR3}.
These Feynman rules represent one gluon emitted from the finite-distance Wilson line in the $n^\mu$ direction, and one or two gluons emitted from the field-strength tensor. 
The respective contribution from diagram $(b)$ is proportional to $C_F$ and already contained in~\eqref{eq:CFZfactor}, whereas the one from $(c)$ is proportional to ($C_F - C_A/2$), and therefore comes with the same integral~\eqref{eq:diagrambCF}. 
Diagram $(g)$, which also needs to be divided by its symmetry factor $2$, is canceled by a piece from diagram $(f)$, and the remaining terms from $(f)$ result again in the expression~\eqref{eq:Zfactorfirstpiece}. 

The Feynman rule that represents one gluon emitted from the Wilson line in the $\bar{n}^\mu$ direction and one from the field-strength tensor is given by the third line of~\eqref{eq:FR2}.
Diagram $(c)$ then contributes the following term to the $Z$-factor,
\begin{align}
    &-\frac{\alpha_s C_A e^{\eps \gamma_E} \mu^{2\eps}}{4 \pi} \delta(\omega -\omega') \Gamma(\eps) \int_0^\infty \! d\ell_- \, \ell_-^{-2\eps} \, \frac{1}{\ell_- - \bar{\Delta}}\Big(\delta(\omega_1 - \omega'_1 + \ell_-) - \delta(\omega_1 - \omega'_1)\Big) \nonumber \\[0.2em]
    = \, &-\frac{\alpha_s C_A}{4 \pi} \delta(\omega -\omega') \frac{1}{\eps} \bigg\{\frac{\theta(\omega'_1-\omega_1)}{\omega'_1-\omega_1 - \bar{\Delta}} - \delta(\omega_1 - \omega'_1) \left(\frac{1}{2\eps} - \ln \frac{-\bar{\Delta}}{\mu} \right) \bigg\} + \mathcal{O}(\eps^0)\,.
\end{align}
Similar to~\eqref{eq:diagrambCF}, the quantity $\bar{\Delta}$, with $\text{Im} \,\bar{\Delta} >0$, is an off-shell regulator for IR divergences, that can be set to zero after expanding in terms of plus-type distributions, which is analogous to~\eqref{eq:diagrambCFdist}. 
The UV-divergent piece of diagram $(f)$ yields the integral
\begin{align}
    &-\frac{\alpha_s C_A}{4 \pi} \delta(\omega -\omega') \frac{1}{\eps} \int_0^{\omega'_1} \frac{d\ell_-}{\ell_- - \bar{\Delta}}\Big(\delta(\omega_1 - \omega'_1 + \ell_-) - \delta(\omega_1 - \omega'_1)\Big) \frac{(\omega'_1 - \ell_-)(2\omega'_1 - \ell_-)}{(\omega'_1)^2}  \nonumber \\[0.2em]
    = \, &-\frac{\alpha_s C_A}{4 \pi} \delta(\omega -\omega') \frac{1}{\eps} \bigg\{-\frac{\omega_1}{(\omega'_1)^2} \big[\theta(\omega_1)\theta(\omega'_1-\omega_1) - \theta(-\omega_1) \theta(\omega_1 - \omega'_1) \big] \\[0.2em]
    &+ \, \frac{2\omega_1 \big[\theta(\omega_1)\theta(\omega'_1-\omega_1) - \theta(-\omega_1)\theta(\omega_1 - \omega'_1)\big]}{\omega'_1(\omega'_1-\omega_1 - \bar{\Delta})} + \delta(\omega_1 - \omega'_1) \left(\frac52 + 2 \ln \frac{\bar{\Delta}}{\bar{\Delta}-\omega_1} \right) \bigg\} \,. \nonumber
\end{align}
Compared to~\eqref{eq:diagrambCFdist}, the additional constraint $\theta(\pm \omega_1)$ in the first term of the last line allows to expand this piece in standard plus-distributions, after which we can set $\bar{\Delta} \to 0$. 
Diagram $(f)$ then results in
\begin{align}
    &-\frac{\alpha_s C_A}{4 \pi} \delta(\omega -\omega') \frac{1}{\eps} \bigg\{-\frac{\omega_1}{(\omega'_1)^2} \big[\theta(\omega_1)\theta(\omega'_1-\omega_1) - \theta(-\omega_1) \theta(\omega_1 - \omega'_1) \big] \nonumber \\[0.2em]
    &+ \, 2\theta(\omega_1) \omega_1 \left[ \frac{\theta(\omega'_1 - \omega_1)}{\omega'_1 (\omega'_1 - \omega_1)} \right]_+ 
    + 2\theta(-\omega_1) \omega_1 \left[ \frac{\theta(\omega_1 - \omega'_1)}{\omega'_1 (\omega_1 - \omega'_1)} \right]_+
    + \frac52 \, \delta(\omega_1 - \omega'_1) \bigg\} \,.
\end{align}

The contribution from diagram $(b)$ turns out to be more subtle.
Carrying out the integration over the $\ell_+$-component is trivial due to the appearing $\delta$-function.
Performing additionally the integration over the perpendicular components yields for the UV-divergent piece
\begin{align}
\label{eq:mixingtermdiagb}
 &-\frac{\alpha_s C_A}{4\pi \eps} \frac{1}{2\pi i} \int \! d\ell_- \, \frac{1}{\omega - \omega' + \ell_- + \delta} \frac{1}{\ell_- - \bar{\Delta}} \Big(\delta(\omega_1 - \omega'_1 + \ell_-) - \delta(\omega_1 - \omega'_1)\Big) \nonumber \\[0.2em]
 = \, &- \frac{\alpha_s C_A}{4\pi \eps} \Big( \frac{1}{2\pi i} \frac{1}{\omega - \omega' + \omega'_1 - \omega_1 + i0} \frac{1}{\omega'_1 - \omega_1 - i0} - \delta(\omega_1 - \omega'_1) \frac{1}{\omega - \omega' + i0}\Big) \,,
\end{align}
with $\delta = 2 vk + i0$ an off-shellness for the residual momentum of the external heavy-quark line.
However, note the important fact that no $\theta$-functions appear in this expression, and the only relevant piece of the IR regulators is their $i0$ prescriptions that dictate how the pole is shifted to the complex plane. 
The equal sign in~\eqref{eq:mixingtermdiagb} has to be understood in that sense. 
In our convention, all IR regulators have a $+i0$ prescription, which matches the Feynman propagators of the corresponding QCD diagrams. 
The second term in~\eqref{eq:mixingtermdiagb} can be further expressed in terms of the $H_\pm$ distributions, defined in~\eqref{eq:FandGdistributions} and~\eqref{eq:H_distributions}, via
\begin{equation}
\label{eq:Hpmidentity}
    \frac{1}{\omega - \omega' + i0} = H_+(\omega,\omega') - H_-(\omega,\omega') - 2 \pi i \delta(\omega - \omega') \,.
\end{equation}
For reasons that become clear later, we do not further manipulate the first term in~\eqref{eq:mixingtermdiagb} at this point.

The last missing contribution arises from diagram $(g)$ using the second term in the second line of the Feynman rule in~\eqref{eq:FR3}.
This contribution describes the contraction of one gluon field from the Wilson line in the $n^\mu$ direction with a gluon field from the $\bar{n}^\mu$ Wilson line. 
The gluon-field from the field-strength tensor is contracted with the external state.
Note that, because the field-strength tensor is contracted with an $\bar{n}_\beta$ in~\eqref{eq:g17def}, there is no non-vanishing contraction of two gluons from $G_s^{\alpha\beta}$ and one gluon from the $\bar{n}^\mu$ Wilson line. 
After performing the integral over the perpendicular components one finds
\begin{align}
\label{eq:ZfactorCAdiagramgnnbarWL}
    &\frac{\alpha_s C_A}{4\pi} \Gamma(\eps) e^{\eps \gamma_E} \mu^{2\eps} \frac{1}{2\pi i} \int \! d\ell_+ \int \! d \ell_- \, (-\ell_+ \ell_- - i0)^{-\eps} \nonumber \\[0.2em]
    &\qquad \qquad \times \frac{1}{\ell_+ + \Delta}\Big(\delta(\omega - \omega' - \ell_+) - \delta(\omega - \omega')\Big) \, \frac{1}{\ell_- - \bar{\Delta}}\Big(\delta(\omega_1 - \omega'_1 + \ell_-) - \delta(\omega_1 - \omega'_1)\Big) \nonumber \\[0.2em]
= & \, -\frac{\alpha_s C_A}{4\pi} \frac{1}{\eps} 
\, \bigg\{ 
\frac{1}{2\pi i} \frac{1}{\omega' - \omega - i0} \frac{1}{\omega'_1 - \omega_1 - i0} - \left(\frac{1}{\eps} + i \pi\right) \delta(\omega - \omega') \delta(\omega_1 - \omega'_1) \nonumber \\
&\qquad\qquad\quad + \delta(\omega_1 - \omega'_1) \left(\frac{\theta(\omega - \omega')}{\omega - \omega' + \Delta} + \delta(\omega - \omega') \ln \frac{\Delta}{\mu} \right) \nonumber \\
&\qquad\qquad\quad + \delta(\omega - \omega') \left(\frac{\theta(\omega_1 - \omega'_1)}{\omega_1 - \omega'_1 + \bar{\Delta}} + \delta(\omega_1 - \omega'_1) \ln \frac{\bar{\Delta}}{\mu} \right)
\bigg\} + \mathcal{O}(\eps^0)\,.
\end{align}
The expression of the last two lines in terms of plus-type distributions reads
\begin{align}
    &\frac{\theta(\omega - \omega')}{\omega - \omega' + \Delta} + \delta(\omega - \omega') \ln \frac{\Delta}{\mu} \nonumber \\
    = \, &\theta(\omega) \left[ \frac{\theta(\omega-\omega')}{\omega-\omega'} \right]_\oplus 
    +\theta(-\omega) \left[\frac{\theta(\omega-\omega')}{\omega'(\omega-\omega')}\right]_+ \omega'
    -\frac12 \ln \frac{\mu^2}{\omega^2} \delta(\omega-\omega') \,,
\end{align}
whereas for the first term in~\eqref{eq:ZfactorCAdiagramgnnbarWL} one could use the identity~\eqref{eq:Hpmidentity}. 
However, we again prefer to not manipulate this term. 

Summing up all pieces, and including the $\mathcal{O}(\alpha_s)$ coupling and gluon-field renormalisation factors
\begin{equation}
  Z_{g_s} \sqrt{Z_3} - 1 = -\frac{\alpha_s C_A}{4\pi \eps} + \mathcal{O}(\alpha_s^2) \,, 
\end{equation}
one finally obtains for the non-Abelian contributions to the $Z$-factor
\begin{align}
\label{eq:CAZfactor}
    Z_{17}^{(1)}(\omega,\omega_1,\omega',\omega'_1;\mu)\Big\vert_{C_A} = \, &\frac{\alpha_s C_A}{4\pi} \delta(\omega - \omega') 
    \bigg\{ \left( \frac{1}{\eps^2} + \frac{1}{\eps}\ln \frac{\mu^2}{\omega_1^2}\right) \delta(\omega_1-\omega'_1) \nonumber \\[0.2em]
    + \, &\frac{2}{\eps} \frac{\omega_1}{(\omega'_1)^2} \big[ \theta(\omega_1)\theta(\omega'_1 - \omega_1) - \theta(-\omega_1) \theta(\omega_1 - \omega'_1) \big] - \frac{1}{\eps} \, \text{Re} \, \mathbf{H}(\omega_1,\omega'_1) \bigg\} \nonumber \\[0.2em]
    + \, &\Delta Z_{17}(\omega,\omega_1,\omega',\omega'_1;\mu) \,,
\end{align}
where we have defined $\mathbf{H}(\omega_1,\omega'_1) = H_+(\omega_1,\omega'_1) + H_-(\omega_1,\omega'_1)$.
The quantity $\Delta Z_{17}$ contains the terms which have non-trivial structures in both sets of arguments $(\omega,\omega')$ and $(\omega_1,\omega'_1)$ simultaneously,
\begin{align}
\label{eq:DeltaZ17}
    \Delta Z_{17}(\omega,\omega_1,\omega',\omega'_1;\mu) 
    = \frac{\alpha_s C_A}{4\pi \eps} \frac{1}{2\pi i} \frac{1}{\omega'_1 - \omega_1 - i0} \Big(\frac{1}{\omega' - \omega + \omega_1 - \omega'_1 - i0}
    - \frac{1}{\omega' - \omega - i0} \Big) \,.
\end{align}
Interestingly, besides these pieces the dependence on $\omega$ in~\eqref{eq:CAZfactor} is purely local, as the distributions exactly cancel in sum of all diagrams. 
The complete one-loop $Z$-factor of the operator $\widetilde{\mathcal{O}}_{17}$ is then given by
\begin{equation}
    Z_{17} = \delta(\omega - \omega') \delta(\omega_1 - \omega'_1) + Z_{17}^{(1)}\Big\vert_{C_F} + Z_{17}^{(1)}\Big\vert_{C_A} + \mathcal{O}(\alpha_s^2) \,,
\end{equation}
where we have suppressed the arguments for brevity.

%----------------------------------------------
\subsection{Cut diagrams and the relevance of \texorpdfstring{$\Delta Z_{17}$}{DZ17}}
\label{subsec:relevanceofDeltaZ}
The appearance of structures like~\eqref{eq:DeltaZ17} in the UV-poles of the soft HQET operator $\mathcal{O}_{17}$ seems problematic.
In the effective theory, the two collinear directions are decoupled, and each hard-collinear sector is described by a jet function that only depends either on $\omega$ (the soft momentum component associated with the $n^\mu$ direction) or $\omega_1$ (the soft momentum component associated with the $\bar{n}^\mu$ direction). 
 
At $\mathcal{O}(\alpha_s)$, all divergences in $\eps$ must cancel in the \empty{sum} of the hard, hard-collinear, anti-hard-collinear and soft loops, where the hard functions do not depend on the soft momenta. 
This would indicate that the $Z$-factor of the operator $\mathcal{O}_{17}$
should be of the form
\begin{equation}
    Z^{(1)}(\omega,\omega_1,\omega',\omega'_1;\mu) = \delta(\omega_1 - \omega'_1) Z_n^{(1)}(\omega,\omega';\mu) + \delta(\omega - \omega') Z_{\bar{n}}^{(1)}(\omega_1,\omega'_1;\mu) \,,
\end{equation}
i.e.~the dependence on the two light-cone momentum projections decouples.
The terms in~\eqref{eq:DeltaZ17} obviously have a different form, and might suggest the presence of other long-range interactions between the two collinear sectors which superficially would violate soft-collinear factorisation. 
However, in the following we will argue that both terms in $\Delta Z_{17}$ are irrelevant for the decay rate, thus, our results agree with factorisation in terms of hard, 
(anti-)hard-collinear and soft modes. 
First, we show that $\Delta Z_{17}$ vanishes in the convolution with the jet functions on the level of time-ordered products. Second, we derive that $\Delta  Z_{17}$ vanishes when the cut is taken into account using cutting rules.

Let us recall that the UV-singularities in~\eqref{eq:CFZfactorfinalnocutoff} and~\eqref{eq:CAZfactor} were calculated using standard Feynman rules from time-ordered products  in a $B$-meson forward matrix element.
In other words, we have not yet applied the necessary cuts that contribute to the $\bar{B} \to X_s \gamma$ decay rate. 
In that case, we will now argue that the problematic terms~\eqref{eq:DeltaZ17} vanish when convoluted with the jet functions due to the location of singularities in the complex plane.
A crucial property for this argument is that the terms in $\Delta Z_{17}$ do not contain any integration constraints from $\theta$-functions, i.e.~in any matrix element of $\widetilde{\mathcal{O}}_{17}$ all light-cone momenta are unrestricted, and hence integrated over the entire real axis in factorisation theorems.\footnote{We argue in appendix~\ref{app:support} that imposing a restricted support with cut-offs leads to inconsistencies with the local limit of the operator.}
Due to the directed energy flow, the relevant (time-ordered) jet functions for the forward matrix element have singularities or branch cuts only in the lower half-plane, i.e.~they depend on $\omega + i0$ and $\omega_1 + i0$, respectively.
At leading order, for example, the jet function in the $n^\mu$ direction is simply given by the inverse hard-collinear propagator $\sim (\omega + i0)^{-1}$, and the same is true for the jet function in the $\bar{n}^\mu$ direction in the massless-quark limit $m_u = 0$. 
The convolution of $\Delta Z_{17}$ with these jet functions can be performed with contour methods, and because the singularities of $\Delta Z_{17}$ in~\eqref{eq:DeltaZ17} are located in the same half-plane, the convolution integrals vanish.
Both terms in the expression~\eqref{eq:DeltaZ17} vanish upon integration in $\omega$, while the $\omega_1$ integration would only eliminate the second term.

We have explicitly checked that $\Delta Z_{17}$ is irrelevant for the cancellation of all $1/\eps$ singularities between hard, (anti-)hard-collinear and soft loops at next-to-leading order.
To do so, we have also computed all $1/\eps$ singularities of the two-loop $\bar{n}$-jet function in the massless case, which will be discussed in a separate publication~\cite{Bartocci25}.
Note that for this cross-check to hold it is important that the limit $m_c \to m_u = 0$ is smooth.  

An alternative point-of-view provides a discussion in position space~\cite{Benzke:2010js}.
The large energy in the $n^\mu$ direction flows from left to right in the Feynman diagram in Fig.~\ref{fig:LO_diagram}. Hence, \emph{after} convolution with the (time-ordered) jet functions, the heavy-quark field with position argument $(tn)$ is evaluated at a later time than the one at space-time point $0$.
This is ensured by the relation
\begin{equation}
    \int \! d\omega \, \frac{e^{-i \omega t}}{\omega + p_+ + i0} = - 2 \pi i e^{i t p_+} \theta(t) \,,
\end{equation}
where $p_+$ is the small component of the hard-collinear momentum.
Indeed, the field $\bar{h}_v(tn)$ appears to the left of $h_v(0)$ in~\eqref{eq:Q17} such that the fields are correctly time-ordered. 
Contrarily, after a Fourier transform to position space, the terms in $\Delta Z_{17}$ are non-zero only for $t<0$, and thus do not contribute.
It would be interesting to investigate the UV singularities of the operator directly in position space, as was done for the soft-quark function in $gg \to h$ at next-to-leading power in~\cite{Beneke:2024cpq}.

Recall that the relevant integrals that contribute to the $\bar{B} \to X_s \gamma$ decay rate contain cuts.
In particular, at leading order, the $n$-jet function is given by the discontinuity of the hard-collinear $s$-quark propagator, whereas the $\bar{n}$-jet function is defined at the amplitude level and arises from the anti-hard-collinear momentum configuration in the quark-loop in Fig.~\ref{fig:LO_diagram}. 
While the potentially problematic terms vanish after convolution with the time-ordered propagators, they certainly do not when convoluted with the physically relevant $n$-jet function. 
However, it turns out that both terms in $\Delta Z_{17}$ arise from soft gluons connecting the amplitude with the complex conjugate amplitude. 
Since taking the cut enforces $t>0$ from the positive-energy constraint, a definition of the shape function $g_{17}(\omega,\omega_1;\mu)$ that contains the relevant cut-propagators would directly eliminate $\Delta Z_{17}$. 
In summary, the contribution $\Delta Z_{17}$, which is non-zero only for $t<0$, is irrelevant in both cases, since either taking the cut or the convolution with the time-ordered jet function enforces $t>0$.

As already mentioned above, the $\bar{B} \to X_s \gamma$ decay rate cannot be expressed as the imaginary part of a $B$-meson forward matrix element of time-ordered operators because not all possible cuts also contribute to the $b \to s \gamma$ decay. Thus, a precise prescription is needed how to implement the restricted cuts when radiative corrections to the process are taken into account.
A path-integral method to evaluate such cut diagrams is given by the Keldysh formalism~\cite{Schwinger:1960qe,Keldysh:1964ud}, see also~\cite{Becher:2007ty} for a concise summary.
Here one introduces fields with a subscript ``$+$'' that belong to the amplitude and are evaluated with standard HQET Feynman rules from time-ordered products. 
On the other hand, fields with a subscript ``$-$'' belong to the complex conjugate amplitude and are evaluated with complex-conjugate Feynman rules from anti-time-ordered products.
The Feynman rule for the contraction of two fields with different indices, i.e. a propagator that connects the amplitude with the complex conjugate amplitude, is evaluated using the on-shell condition on the right-hand side in~\eqref{eq:cutpropagator}.
On a diagrammatic level, these rules precisely correspond to the cutting-rules explained in the beginning of this section. 
However, since the shape function $g_{17}$ is defined by a hadronic matrix element, a non-perturbative definition at the level of the path integral is required.
The soft operator~\eqref{eq:Q17} that defines the shape function $g_{17}$ should, thus, be written as
\begin{equation}
\label{eq:Q17Keldysh}
    \mathcal{O}_{17}(t,r) \to \big(\bar h_v S_n\big)_-(tn)\,
    \nbslash \big(S_n^\dagger S_{\bar n}\big)_+(0)\,
    i\gamma_\alpha^\perp\bar n_\beta\,
    \big(S_{\bar n}^\dagger\,g_s G_s^{\alpha\beta} S_{\bar n} 
    \big)_+(r\bar n)\,
    \big(S_{\bar n}^\dagger h_v\big)_+(0) \,.
\end{equation}
This prescription does not affect the Feynman rules for the effective vertex provided in Section~\ref{sec:renormalization}, but the use of cut propagators from~\eqref{eq:cutpropagator} introduces some relevant differences in a few diagrams.
Indeed, evaluating all diagrams in Fig.~\ref{fig:g17_diagrams} with these cutting rules leaves most of the results from the previous two subsections unchanged.
In particular the Abelian part of the $Z$-factor, as well as the sum of diagrams $(a)$ and $(c)$-$(f)$ remain unaffected.
Notable differences are that the contribution~\eqref{eq:mixingtermdiagb} becomes UV-finite, and second, the expression in~\eqref{eq:ZfactorCAdiagramgnnbarWL} becomes
\begin{align}
-\frac{\alpha_s C_A}{4\pi} \frac{1}{\eps} 
\, \bigg\{ 
&- \frac{1}{\eps} \delta(\omega - \omega') \delta(\omega_1 - \omega'_1) 
+ \delta(\omega_1 - \omega'_1) \left(\frac{\theta(\omega' - \omega)}{\omega' - \omega - \Delta} + \delta(\omega - \omega') \ln \frac{-\Delta}{\mu}  \right) \nonumber \\[0.2em]
&+ \delta(\omega - \omega') \left(\frac{\theta(\omega_1 - \omega'_1)}{\omega_1 - \omega'_1 + \bar{\Delta}} + \delta(\omega_1 - \omega'_1) \ln \frac{\bar{\Delta}}{\mu} \right)
\bigg\} + \mathcal{O}(\eps^0)\,.
\end{align}
This is indeed the sum of~\eqref{eq:mixingtermdiagb} and~\eqref{eq:ZfactorCAdiagramgnnbarWL} apart from the two terms in $\Delta Z_{17}$.
In other words, one finds the same soft UV poles when computing time-ordered diagrams and dropping irrelevant terms that vanish after convolution with the jet functions, or by directly computing cut diagrams.
The reason for this is likely the absence of imaginary parts in the non-vanishing soft contributions, as in that case the cuts in the respective diagrams for the decay rate only affect the $s$-quark propagator, which belongs to the $n$-jet function.
The shape function $g_{17}$ that appears in the factorisation theorem for the $\bar{B} \to X_s \gamma$ decay rate, and is convoluted with the discontinuity of the $n$-jet function, is given by the operator~\eqref{eq:Q17Keldysh} including the cuts through the Keldysh formalism.
We will adopt this prescription in the following.
Evaluating the matrix element with time-ordered products, on the other hand, would be the relevant soft function that appears in the factorisation of the $B$-meson forward matrix element of time-ordered weak Hamiltonians, i.e. before the restricted cuts are taken.

Interestingly, such irrelevant contributions also arise for soft functions defined at the amplitude level, for example in exclusive $B$ decays.
In this case indeed only ordinary (time-ordered) Feynman diagrams contribute.
We will discuss an example from the literature~\cite{Huang:2023jdu} in greater detail in section~\ref{sec:exclusive}.

%----------------------------------------------
\section{Analytic solution to the renormalisation-group equation}
\label{subsec:solution}
In this section we present an analytic solution to the RG equation for the subleading shape function 
$g_{17}(\omega,\omega_1;\mu)$ as defined using time-ordered and anti-time-ordered fields in~\eqref{eq:Q17Keldysh},
\begin{equation}
\frac{d}{d \ln \mu} \, g_{17}(\omega, \omega_1; \mu)
= - \int \! d \omega' \int \! d \omega'_1 \, \gamma_{17} (\omega, \omega_1, \omega', \omega'_1; \mu) \,  g_{17}(\omega', \omega'_1; \mu) \,,
\label{eq:RGE_general}
\end{equation}
in momentum space.
To do so, let us first recall that the anomalous dimension which governs the scale evolution of $g_{17}(\omega,\omega_1;\mu$) is defined as
\begin{equation}
\gamma_{17}(\omega,\omega_1,\omega',\omega'_1;\mu)
= - \int \!  d \hat{\omega}  \, \int \! d \hat{\omega}_1 \,
\frac{d Z_{17}(\omega, \omega_1, \hat{\omega}, \hat{\omega}_1; \mu)}{d \ln \mu} \, 
Z_{17}^{-1}(\hat{\omega}, \hat{\omega}_1, \omega', \omega'_1; \mu) \,,
\end{equation}
with the $Z$-factor from~\eqref{eq:CFZfactorfinalnocutoff} and~\eqref{eq:CAZfactor} 
without $\Delta Z_{17}$. 
At $\mathcal{O}(\alpha_s)$, we can decompose the anomalous dimension as a sum of two pieces, each of which is associated with only one of the two light-cone directions,
\begin{align}
\label{eq:gamma17}
 \gamma_{17}(\omega,\omega_1,\omega',\omega'_1;\mu)
 = \frac{\alpha_s}{\pi} \Big\{ C_F \, \delta(\omega_1 - \omega'_1) \gamma_{n}(\omega,\omega';\mu)
+ \frac{C_A}{2} \, \delta(\omega - \omega') \gamma_{\bar{n}}(\omega_1,\omega'_1;\mu) \Big\} \,. 
\end{align}
The Abelian part
\begin{equation}
\label{eq:gamman}
    \gamma_{n}(\omega,\omega';\mu) = \left(\ln \frac{\mu^2}{\omega^2} - 1\right) \delta(\omega-\omega')
    - 2\theta(\omega) \left[\frac{\theta(\omega'-\omega)}{\omega'(\omega'-\omega)}\right]_+ \omega' 
    - 2\theta(-\omega) \left[\frac{\theta(\omega'-\omega)}{\omega'-\omega}\right]_\ominus
\end{equation}
only acts on the soft variables associated with the collinear direction, and the non-Abelian part
\begin{equation}
\label{eq:gammanbar}
    \gamma_{\bar{n}}(\omega_1,\omega'_1;\mu) = \ln \frac{\mu^2}{\omega_1^2} \, \delta(\omega_1-\omega'_1) - \text{Re} \, \mathbf{H}(\omega_1,\omega'_1) 
    + \frac{2 \omega_1}{(\omega'_1)^2} \big[ \theta(\omega_1)\theta(\omega'_1 - \omega_1) - \theta(-\omega_1) \theta(\omega_1 - \omega'_1) \big]
\end{equation}
only acts on the soft variables associated with the anti-collinear direction.
Because the dependence on the two momentum variables $\omega$ and $\omega_1$ factorizes, the RG equation~\eqref{eq:RGE_general} can be separated in two independent equations which we solve consecutively.

%----------------------------------------------
\subsection{Abelian part}
\label{subsec:solutionAbelianpart}
Our strategy to solve the integro-differential evolution equation closely follows the discussion from~\cite{Beneke:2022msp}. 
The RG equation in the Abelian limit is identical to the one of the leading shape function.
To introduce some notation and concepts for the later sections, we now re-derive its well-known solution~\cite{Bosch:2004th}.   

To do so, we first divide the support of the shape function in the variable $\omega$ into two branches,
\begin{equation}
\label{eq:supportsplit}
    g_{17}(\omega,\omega_1;\mu) = \theta(\omega) g_{17}^{>}(\omega,\omega_1;\mu) + \theta(-\omega) g_{17}^{<} (\omega,\omega_1;\mu) \,,
\end{equation}
after which we perform a Mellin transformation that translates the distributions into ordinary functions or derivatives in the conjugate variable~\cite{Bell:2013tfa,Braun:2014owa, Lee:2005gza}.
The Mellin transform is defined separately for positive values of $\omega$ via
\begin{align}
    \label{eq:Laplace_positive}
    \tilde{g}_{17}^>(\eta,\omega_1;\mu) &= \int_0^\infty \frac{d\omega}{\omega} \left(\frac{\mu}{\omega}\right)^{\!\eta} g_{17}^>(\omega,\omega_1;\mu) \; , \nonumber\\ 
    g_{17}^>(\omega,\omega_1;\mu) &= \int_{c-i\infty}^{c+i\infty} \frac{d\eta}{2\pi i} \left( \frac{\mu}{\omega}\right)^{\! -\eta} \tilde{g}_{17}^>(\eta,\omega_1;\mu) \,,
\end{align}
and for negative $\omega$ by
\begin{align}
\label{eq:Laplace_negative}
    \tilde{g}_{17}^<(\eta,\omega_1;\mu) &= \int^{\infty}_0 \frac{d\omega}{\omega} \left( \frac{\mu}{\omega} \right)^{\!\eta} g_{17}^<(-\omega,\omega_1;\mu) \,, \nonumber\\
    g_{17}^<(-\omega,\omega_1;\mu)& = \int_{c-i\infty}^{c+i\infty} \frac{d\eta}{2\pi i} \left(\frac{\mu}{\omega} \right)^{\!-\eta} \tilde{g}_{17}^<(\eta,\omega_1;\mu) \, .
\end{align}
The expressions for the different types of distributions convoluted with pure powers in $\omega$ are collected in Appendix B of~\cite{Beneke:2022msp}, and we do not repeat them here.
In Mellin space, the Abelian part of the evolution equation turns into the following coupled system of differential equations
\begin{align}
    \left( \frac{d}{d\ln \mu} -\eta \right) \tilde{g}_{17}^>(\eta,\omega_1;\mu) = \,&\frac{\alpha_s C_F}{\pi} \Big[ -2H_{-1-\eta} -2\partial_\eta + 1 \Big]\tilde{g}_{17}^>(\eta,\omega_1;\mu)\, ,  \label{eq:wRGE}\\[0.2em]
    \left( \frac{d}{d\ln \mu} -\eta \right) \tilde{g}_{17}^<(\eta,\omega_1;\mu) = \, &\frac{\alpha_s C_F}{\pi} \Big[ - 2H_{\eta} -2\partial_\eta + 1 \Big]\tilde{g}_{17}^<(\eta,\omega_1;\mu) \nonumber \\
    &+ \frac{\alpha_s C_F}{\pi} 2\Gamma(-\eta) \Gamma(1+\eta)  \tilde{g}_{17}^>(\eta,\omega_1;\mu)\label{eq:wRGE2} \,,
\end{align}
which hold for $-1<\text{Re}(\eta)<0$, and $H_\eta$ is the Harmonic number function.
This system can be diagonalized by choosing $\tilde{g}_{17}^>(\eta,\omega_1;\mu)$ and the linear combination
\begin{equation}
\label{eq:diagonal_function_w}
\tilde{g}_{17}(\eta,\omega_1;\mu)
\equiv \int_{-\infty}^{+\infty} \! \frac{d\omega}{-\omega-i0} \left( \frac{\mu}{-\omega-i0} \right)^\eta g_{17}(\omega,\omega_1;\mu) 
= \tilde{g}_{17}^<(\eta,\omega_1;\mu) - \tilde{g}_{17}^>(\eta,\omega_1;\mu)e^{i\pi\eta} \,,
\end{equation}
as the two independent functions. 
The latter fulfills
\begin{equation}
\label{eq:diagonal_RGE_w}
   \left( \frac{d}{d\ln \mu} -\eta \right) \tilde{g}_{17}(\eta,\omega_1;\mu)  = \frac{\alpha_s C_F}{\pi} \Big[ -2H_{\eta} -2\partial_\eta + 1 \Big] \tilde{g}_{17}(\eta,\omega_1;\mu) \,,
\end{equation}
which can be easily verified using the identity $\Gamma(-\eta) \Gamma(1+\eta) = e^{i \pi \eta} (i\pi + H_\eta - H_{-1-\eta})$.
We note that the integral in~\eqref{eq:diagonal_function_w} does not have an inverse transformation and should therefore not be considered as an integral transform. 

Instead of solving~\eqref{eq:wRGE} and~\eqref{eq:wRGE2}, it is simpler to solve~\eqref{eq:wRGE} and~\eqref{eq:diagonal_RGE_w} and then use~\eqref{eq:diagonal_function_w} to obtain $\tilde{g}_{17}^<(\eta,\omega_1;\mu)$.
The two solutions to the diagonal equations read
\begin{align} 
    \tilde{g}_{17}^>(\eta,\omega_1;\mu) &= e^{2V + 2\gamma_E a} \left( \frac{\mu}{\mu_0} \right)^{\!\!\eta} \frac{\Gamma(-\eta)}{\Gamma(-\eta-2a)} \,\tilde{g}_{17}^>(\eta+2a,\omega_1;\mu_0) \; , \label{eq:wsol1} \\[0.2em]
    \tilde{g}_{17}(\eta,\omega_1;\mu) &= e^{2V + 2\gamma_E a} \left( \frac{\mu}{\mu_0} \right)^{\!\!\eta} \frac{\Gamma(1+\eta+2a)}{\Gamma(1+\eta)} \,\tilde{g}_{17}(\eta+2a,\omega_1;\mu_0) \,, \label{eq:wsol2}
\end{align}
with the evolution factors
\begin{align}
\label{eq:VaQCDCF}
    V(\mu,\mu_0)  =&-\int_{\mu_0}^\mu \frac{d\mu'}{\mu'} \frac{\alpha_s(\mu') C_F }{\pi} \bigg[  \ln \frac{\mu'}{\mu_0} -\frac12  \bigg] \; , \nonumber \\ 
    a(\mu,\mu_0) = &-\int_{\mu_0}^\mu \frac{d\mu'}{\mu'} \frac{\alpha_s(\mu')C_F}{\pi} \,,
\end{align}
whose dependence on $\mu$ and $\mu_0$ has been omitted in~\eqref{eq:wsol1} and~\eqref{eq:wsol2} for brevity.
Further, in the integrand we can identify the factor $\gamma_{\rm cusp} = \alpha_s C_F/\pi + \mathcal{O}(\alpha_s^2)$ with the leading term of the universal cusp anomalous dimension.

After performing the inverse transformations~\eqref{eq:Laplace_positive} and \eqref{eq:Laplace_negative}, one finds the momentum-space solution for the $\omega > 0$ branch,
\begin{equation}
\label{momspaceCFsolution_pos}
 g_{17}^{>}(\omega,\omega_1;\mu) = \frac{e^{2V + 2\gamma_E a}}{\Gamma(-2a)} \int_\omega^\infty \! \frac{d\omega'}{\omega' - \omega} \, \left(\frac{\mu_0}{\omega' - \omega}\right)^{\!2a} g_{17}^>(\omega',\omega_1;\mu_0) \,,
 \end{equation}
and for the $\omega < 0$ branch,
\begin{align}\label{momspaceCFsolution_neg}
 g_{17}^{<}(\omega,\omega_1;\mu) = \frac{e^{2V + 2\gamma_E a}}{\Gamma(-2a)} \bigg[&\int_0^\infty \! \frac{d\omega'}{\omega' - \omega} \, \left(\frac{\mu_0}{\omega' - \omega}\right)^{\!2a} g_{17}^>(\omega',\omega_1;\mu_0) \nonumber \\[0.2em]
 + &\int_\omega^0 \! \frac{d\omega'}{\omega' - \omega} \, \left(\frac{\mu_0}{\omega' - \omega}\right)^{\!2a} g_{17}^<(\omega',\omega_1;\mu_0)\bigg]\,.
\end{align}
Combining the two then simply yields
\begin{equation}
\label{eq:momspaceCFsolution}
    g_{17}(\omega,\omega_1;\mu) = \frac{e^{2V + 2\gamma_E a}}{\Gamma(-2a)} \int_\omega^\infty \! \frac{d\omega'}{\omega' - \omega} \, \left(\frac{\mu_0}{\omega' - \omega}\right)^{\!2a} g_{17}(\omega',\omega_1;\mu_0) \,.
\end{equation}
This result trivially reproduces the one presented in~\cite{Bosch:2004th} for the leading shape function, once a restricted support $\omega' \in (-\infty, \bar{\Lambda}]$ is assumed for the initial condition $g_{17}(\omega',\omega_1;\mu_0)$.
Recall that this is only consistent with renormalisation if we adopt the definition of $g_{17}(\omega,\omega_1;\mu)$ in terms of time-ordered and anti-time-ordered fields in~\eqref{eq:Q17Keldysh}.
From now on we will explicitly include the upper cut-off in $\omega$.

%----------------------------------------------
\subsection{Non-Abelian part}
\label{subsec:solutionnonAbelianpart}
We now move on to the non-Abelian part of the evolution equation, and adopt the same procedure as discussed before.
First, we perform a similar split into a positive-support and negative-support branch,
\begin{equation}
\label{eq:supportsplitCA}
    g_{17}(\omega,\omega_1;\mu) = \theta(\omega_1) g_{17}^{>}(\omega,\omega_1;\mu) + \theta(-\omega_1) g_{17}^{<} (\omega,\omega_1;\mu) \,.
\end{equation}
For compactness of the notation we use the same symbols as in~\eqref{eq:supportsplit}.
The variable to which the superscripts $>$ and $<$ refer is always clear from the context: 
In this subsection~\ref{subsec:solutionnonAbelianpart} they always refer to $\omega_1$, whereas in the previous subsection~\ref{subsec:solutionAbelianpart} they refer to $\omega$.

Transforming the distributions from~\eqref{eq:gammanbar} to Mellin space yields the differential equations
\begin{align}
    \label{eq:w1RGE}
    &\left(\frac{d}{d\ln \mu} -\eta_1 \right) \tilde{g}_{17}^>(\omega,\eta_1;\mu) \\[0.2em]
    = &- \frac{\alpha_s C_A}{2\pi} \Big\{ \big[ H_{-1-\eta_1}+H_{\eta_1}+ 2H_{1-\eta_1} + 2\partial_{\eta_1} \big] \, \tilde{g}_{17}^>(\omega,\eta_1;\mu)
    - \Gamma(-\eta_1) \Gamma(1+\eta_1) \,   \tilde{g}_{17}^<(\omega,\eta_1;\mu) \Big\} \,, \nonumber
\end{align}
and
\begin{align}
    \label{eq:w1RGE2}
    &\left(\frac{d}{d\ln \mu} -\eta_1 \right) \tilde{g}_{17}^<(\omega,\eta_1;\mu) \\[0.2em]
    = &- \frac{\alpha_s C_A}{2\pi} \Big\{ \big[ H_{-1-\eta_1}+H_{\eta_1}+ 2H_{1-\eta_1} + 2\partial_{\eta_1} \big] \, \tilde{g}_{17}^<(\omega,\eta_1;\mu)
    - \Gamma(-\eta_1) \Gamma(1+\eta_1) \,   \tilde{g}_{17}^>(\omega,\eta_1;\mu) \Big\} \,, \nonumber
\end{align}
which again hold for $-1<\text{Re}(\eta_1)<0$, where $\eta_1$ is the Mellin-space variable conjugate to $\omega_1$.
Since this system is symmetric, it can be diagonalized by using the sum and difference $\tilde{g}^{\pm}_{17}(\omega,\eta_1;\mu) \equiv \tilde{g}_{17}^>(\omega,\eta_1;\mu) \pm \tilde{g}_{17}^<(\omega,\eta_1;\mu)$ as independent functions.
However, the equations become somewhat simpler if one uses the following two linear combinations,
\begin{align}
\label{eq:diagonal_function_w1}
    \tilde{g}_{17}^{(A)}(\omega,\eta_1;\mu) &= \tilde{g}_{17}^>(\omega,\eta_1;\mu)e^{\frac{i\pi\eta_1}{2}}-\tilde{g}_{17}^<(\omega,\eta_1;\mu)e^{-\frac{i\pi\eta_1}{2}} \,, \nonumber \\[0.2em]
    \tilde{g}_{17}^{(B)}(\omega,\eta_1;\mu) &= \tilde{g}_{17}^<(\omega,\eta_1;\mu)e^{\frac{i\pi\eta_1}{2}}-\tilde{g}_{17}^>(\omega,\eta_1;\mu)e^{-\frac{i\pi\eta_1}{2}} \,.
\end{align}
Interestingly, it then turns out that both functions fulfill the same diagonal RG equation,
\begin{equation}\label{eq:RGECAdiag}
    \left( \frac{d}{d\ln \mu} -\eta_1 \right) \tilde{g}_{17}^{(A,B)}(\omega,\eta_1;\mu)  = -\frac{\alpha_s C_A}{\pi} \bigg[ H_{\eta_1} + H_{1-\eta_1} +\partial_{\eta_1} \bigg]\tilde{g}_{17}^{(A,B)}(\omega,\eta_1;\mu)\,,
\end{equation}
which is solved by
\begin{align} 
    \tilde{g}_{17}^{(A,B)}(\omega,\eta_1;\mu) &= e^{V_1+2\gamma_E a_1} \left( \frac{\mu}{\mu_0} \right)^{\!\!\eta_1} \frac{\Gamma(2-\eta_1)\Gamma(1+\eta_1+a_1)}{\Gamma(2-\eta_1-a_1)\Gamma(1+\eta_1)} \,\tilde{g}_{17}^{(A,B)}(\omega,\eta_1+a_1;\mu_0) \,, 
\end{align}
where now the evolution factors read
\begin{align} \label{eq:VaQCDCA}
    V_1(\mu,\mu_0)  =&-\int_{\mu_0}^\mu \frac{d\mu'}{\mu'} \frac{\alpha_s(\mu') C_A }{\pi}   \ln \frac{\mu'}{\mu_0} \; , \nonumber \\ 
    a_1(\mu,\mu_0) = &-\int_{\mu_0}^\mu \frac{d\mu'}{\mu'} \frac{\alpha_s(\mu')C_A}{\pi} \,.
\end{align}

Solving~\eqref{eq:diagonal_function_w1} for the two branches $\tilde{g}_{17}^{>,<}$ yields
\begin{align}
\label{eq:inv_function_w1}
    \tilde{g}_{17}^{>,<}(\omega,\eta_1;\mu) =&-e^{V_1+2\gamma_E a_1} \left( \frac{\mu}{\mu_0} \right)^{\!\!\eta_1} \frac{\Gamma(-\eta_1)\Gamma(2-\eta_1)\Gamma(1+\eta_1+a_1)}{\Gamma(2-\eta_1-a_1)} \bigg[ \\[0.2em]
&\frac{1}{\pi}\sin\left(\big(\eta_1 + \frac{a_1}{2}\big) \pi\right)\tilde{g}_{17}^{>,<}(\omega,\eta_1+a_1;\mu)
+\frac{1}{\pi}\sin\left(\frac{a_1\pi}{2} \right)\tilde{g}_{17}^{<,>}(\omega,\eta_1+a_1;\mu)\bigg]\,. \nonumber 
\end{align}
After replacing the sine functions by gamma functions via the reflection formula, the inverse Mellin transform can be expressed in terms of Meijer-G functions, which are defined by the complex contour integral 
\begin{equation}
    G^{m,n}_{p,q} \bigg( {\mathbf{a} \atop \mathbf{b} } \bigg| z \bigg) = \int \frac{d\eta}{2\pi i} z^\eta \frac{\prod_{j=1}^{m}\Gamma(b_j-\eta)\prod_{j=1}^{n}\Gamma(1-a_j+\eta)}{\prod_{j=m+1}^{q}\Gamma(1-b_j+\eta) \prod_{j=n+1}^{p}\Gamma(a_j-\eta)} 
\end{equation}
for integer $0\leq m \leq q$ and $0\leq n \leq p$, where $\mathbf{a}=(a_1,\dots,a_p)$ and $\mathbf{b}=(b_1,\dots,b_q)$.
Further details and properties of these functions are e.g. outlined in Appendix D of~\cite{Beneke:2022msp}.
The resulting momentum-space solution reads
\begin{equation} 
    {g}_{17}(\omega,\omega_1;\mu) = \int \frac{d\omega'_1}{\big\vert \omega'_1 \big\vert} \, U_{\bar{n}}^{(17)}(\omega_1,\omega'_1;\mu,\mu_0) g_{17}(\omega,\omega_1^\prime;\mu_0) \,, 
\end{equation}
with the evolution function
\begin{align} 
\label{eq:CAevolutionfunction}
     U_{\bar{n}}^{(17)}(\omega_1,\omega'_1;\mu,\mu_0) 
     = \, &-e^{V_{1}+2\gamma_E a_{1}} \left(\frac{\mu_0}{\vert \omega'_1 \vert}\right)^{\! a_1} \bigg\{ 
     \theta(\tau) G^{1,2}_{3,3} \bigg( 
     \begin{array}{cccc} -1, \!\!  &1, \!\!  &a_1/2 \\  a_1+1, \!\!  &a_1-1, \!\! &a_1/2 \end{array}
     \bigg| \, \tau \bigg) \\[0.2em]
     + \, \frac{1}{2 \pi} \sin&\left(\frac{a_1 \pi}{2} \right)\theta(-\tau) 
     \Gamma(1+a_1)  \Gamma(3+a_1) (-\tau)^{1+a_1} {}_2F_1(1+a_1,3+a_1,3;\tau) \bigg\} \,, \nonumber 
\end{align}
where we defined the ratio $\tau = \omega'_1/\omega_1$.

The appearing Meijer-G function can be reduced to a hypergeometric function on the interval $1>\tau>-1$ via
\begin{align}
    &G^{1,2}_{3,3} \bigg( 
     \begin{array}{cccc} -1, \!\!  &1, \!\!  &a_1/2 \\  a_1+1, \!\!  &a_1-1, \!\! &a_1/2 \end{array}
     \bigg| \, \tau \bigg)\nonumber \\[0.2em]
     = \, &\frac{1}{2\pi} \sin\left(\frac{a_1 \pi}{2}\right)\Gamma(1+a_1)    \Gamma(3+a_1) \tau^{1+a_1} {}_2F_1(1+a_1,3+a_1,3;\tau) \,, 
\end{align}
where the variable $\tau$ is implicitly supplemented with an $+i0$ prescription.
From this identity it follows that the evolution function in~\eqref{eq:CAevolutionfunction} is continuous at $\tau = 0$.

%----------------------------------------------
\subsection{Phenomenological implications}
\label{subsec:pheno}
To summarize, the solution to the RG equation takes the factorized form
\begin{equation} \label{eq:g17_evolution}
    {g}_{17}(\omega,\omega_1;\mu) = \int_\omega^{\Bar{\Lambda}}\frac{d\omega'}{\omega'-\omega} \, U_n^{(17)}(\omega,\omega';\mu,\mu_0) \int_{-\infty}^{\infty} \frac{d\omega'_1}{\big\vert \omega'_1 \big\vert} \, U_{\bar{n}}^{(17)}(\omega_1,\omega'_1;\mu,\mu_0) \, g_{17}(\omega',\omega_1^\prime;\mu_0) \,,
\end{equation}
with the evolution function for the soft momenta associated with the $n^\mu$ light-cone,
\begin{equation}
   U_n^{(17)}(\omega,\omega';\mu,\mu_0) = \frac{e^{2V + 2\gamma_E a}}{\Gamma(-2a)} \, \left(\frac{\mu_0}{\omega'-\omega}\right)^{\!2a}\,,
\end{equation}
and the evolution function for the soft momenta associated with the $\bar{n}^\mu$ light-cone given by~\eqref{eq:CAevolutionfunction}. 
Below we will discuss some consequences of RG evolution relevant in phenomenological applications. 
A serious numerical estimate of the evolution effects, however, is left for future work. 

\paragraph{Existence of convolution integrals:}
The functional behavior of the shape function for $\omega\to \Bar{\Lambda}$ and $\omega\to -\infty$ that follows from its $\mathcal{O}(\alpha_s)$ anomalous dimension is known from the literature~\cite{Bosch:2004th}.
Assuming that the function at the low scale $\mu_0$ falls off like $\omega^{-\xi}$ for $\omega\to -\infty$, one finds that for $\mu > \mu_0$ it falls off as $\sim \omega^{-2a - \text{min}(1,\xi)}$, with $a < 0$~\cite{Bosch:2004th}.
For $\omega \to \Bar{\Lambda}$, on the other hand, assuming that the shape function at the low scale vanishes as $(\Bar{\Lambda}-\omega)^{\xi}$, at higher scales it vanishes as $\sim (\Bar{\Lambda}-\omega)^{-2a+\xi}$.
Inverse moments of the shape function converge at the scale $\mu_0$ as long as $\xi>0$, but typically one would assume $\xi > 1$ such that the normalization integral, i.e. the local limit, exists. 
After evolution to the scale $\mu$, inverse moments then converge as long as $2a > -1$, which is true for realistic scales.
It is well known that the normalization integral as well as positive moments, however, become divergent, because the RG evolution generates a radiative tail that falls off slower than $1/\omega$~\cite{Bosch:2004th}.
Given that the cut-propagators from the jet function impose a lower cut-off on $\omega$, convolution integrals in factorisation theorems converge for all practical purposes.

The asymptotic behavior of $g_{17}(\omega,\omega_1;\mu)$ for $\omega_1 \to \pm \infty$ is determined by the location of singularities of the Mellin-space solution~\eqref{eq:inv_function_w1}. 
Assuming $g_{17}(\omega,\omega_1\to \pm \infty;\mu_0) \sim \omega_1^{-\xi_1}$, its Mellin transform has a singularity at $\eta_1 = -\xi_1$.
We deform the integration contour to enclose all poles and cuts on the left half-plane with respect to $\text{Re}(\eta_1)=c$, where $-1-a_1<c<0$.
The singularities of the initial condition with shifted argument at $\eta_1 = -\xi_1 -a_1$, as well as of the gamma function at $\eta_1 = -1 -a_1$ lead to
\begin{equation}
    g_{17}(\omega,\omega_1\to\pm\infty;\mu) \sim \omega_1^{-a_1-\text{min}(1,\xi_1)} \,.
\end{equation}
For the limit $\omega_1\to 0$, the singularity from $\Gamma(-\eta_1)$ determines that the function approaches a constant. 
After expanding the Mellin-space solutions~\eqref{eq:inv_function_w1} around $\eta_1 = 0$, one finds $\tilde{g}_{17}^>(\omega,\eta_1;\mu) \simeq \tilde{g}_{17}^<(\omega,\eta_1;\mu)$, and it follows that $g_{17}(\omega,\omega_1;\mu)$ is continuous at $\omega_1 = 0$. 
This guarantees that principal-value integrals of the form
\begin{equation}
    \int \frac{d\omega_1}{\omega_1 + i0} \, g_{17}(\omega,\omega_1;\mu) = {\rm P} \! \int \frac{d\omega_1}{\omega_1} \, g_{17}(\omega,\omega_1;\mu) - i \pi \, g_{17}(\omega,0;\mu)
\end{equation}
are also well-defined. 
Hence, also the convolution integrals in $\omega_1$ in factorisation theorems exist for all practical purposes.

\paragraph{Properties of the shape function:}  
Two important properties of the shape function $g_{17}(\omega,\omega_1;\mu)$ were derived in~\cite{Benzke:2010js}. 
First, the function is real-valued, i.e. does not carry any strong phases.
Because the anomalous dimension in~\eqref{eq:gamma17} itself is also real-valued, the function trivially does not develop strong phases through RG evolution. 
Second, the shape function obeys the identity
\begin{equation}
\label{eq:intidentity}
    \int_{-\infty}^{\bar{\Lambda}} \! d\omega \, g_{17}(\omega,\omega_1;\mu_0) 
    = \int_{-\infty}^{\bar{\Lambda}} \! d\omega \, g_{17}(\omega,-\omega_1;\mu_0) \,,
\end{equation}
which becomes ill-defined at scales $\mu>\mu_0$ due to the radiative tail for large $\omega$ discussed in the previous paragraph.

Nevertheless, the anomalous dimension associated with the $\bar{n}^\mu$ sector~\eqref{eq:gammanbar} is anti-symmetric under the exchange $\omega_1 \leftrightarrow - \omega_1$ and $\omega'_1 \leftrightarrow - \omega'_1$.
As a consequence, a symmetric function in $\omega_1$ remains symmetric under RG evolution. 
To see this, assume that at some initial scale  $g_{17}^<(\omega,\omega_1;\mu_0) = g_{17}^>(\omega,-\omega_1;\mu_0)$ holds, where the $>$ and $<$ signs refer to the variable $\omega_1$.
In that case, the solution to the RG equation in the $\bar{n}^\mu$ sector can be expressed as
\begin{equation} 
    g_{17}(\omega,\omega_1;\mu) = \int_0^\infty \frac{d\omega'_1}{\omega'_1} \, U_{\bar{n}}^{\rm sym}(\omega_1,\omega'_1;\mu,\mu_0) \, g_{17}(\omega,\omega_1^\prime;\mu_0) \, ,
    \label{eq:w1solsymm} 
\end{equation}
with
\begin{align} 
\label{eq:CAevolutionfunctionsymm}
     U_{\bar{n}}^{\rm sym}(\omega_1,\omega'_1;\mu,\mu_0) 
     = \, &-e^{V_{1}+2\gamma_E a_{1}} \left(\frac{\mu_0}{\omega'_1 }\right)^{\! a_1} \bigg\{ G^{1,2}_{3,3} \bigg( 
     \begin{array}{cccc} -1, \!\!  &1, \!\!  &a_1/2 \\  a_1+1, \!\!  &a_1-1, \!\! &a_1/2 \end{array}
     \bigg| \, |\tau| \bigg) 
     \\[0.2em] 
     + \frac{1}{2 \pi} &\sin\left(\frac{a_1 \pi}{2} \right) 
     \Gamma(1+a_1)  \Gamma(3+a_1) |\tau|^{1+a_1} {}_2F_1(1+a_1,3+a_1,3; -|\tau|) \bigg\} \,. \nonumber 
\end{align}
Because the absolute value $|\tau| = |\omega'_1/\omega_1|$ appears as an argument of the Meijer-G functions, it is easy to see that $g_{17}^<(\omega,\omega_1;\mu) = g_{17}^>(\omega,-\omega_1;\mu)$ holds at any scale $\mu$.

\paragraph{Numerical solution via discretisation:}
\begin{figure}[t]
    \centering
    \includegraphics[width=0.78\textwidth]{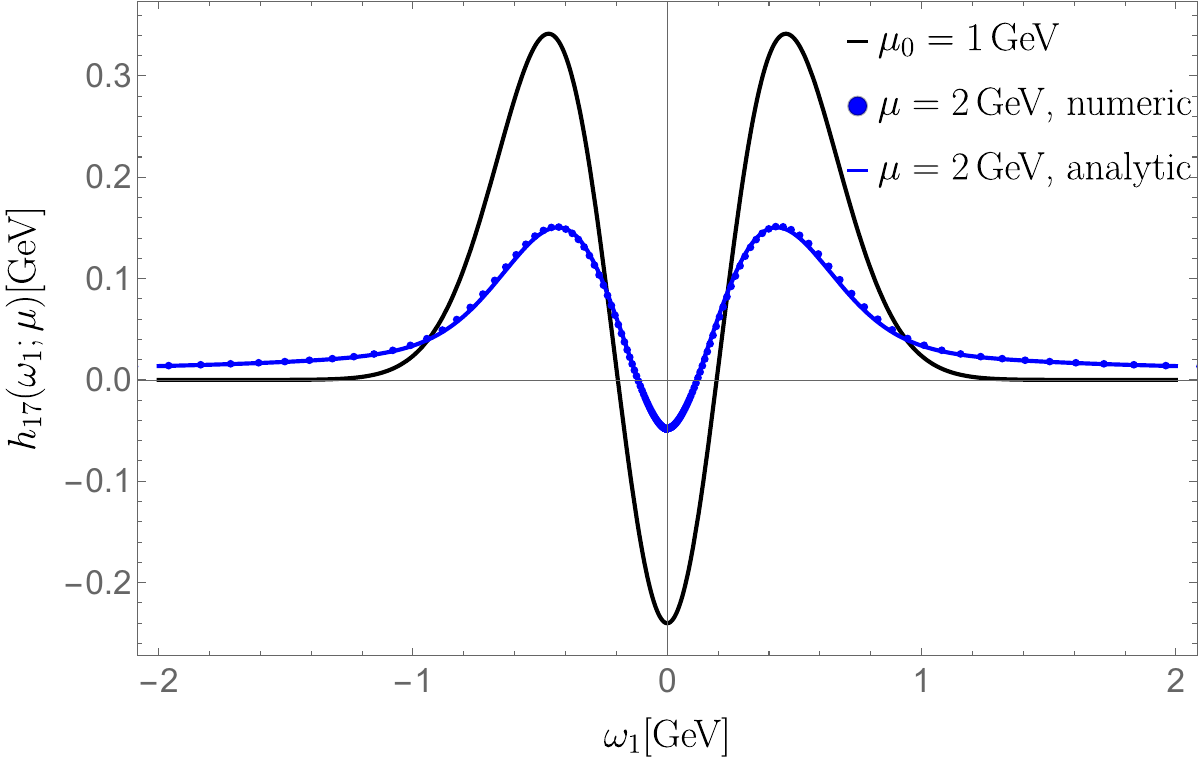}
    \caption{Scale evolution of the function $h_{17}(\omega_1;\mu)$, using the model~\eqref{eq:Hermitemodel} with $n \leq 2$ Hermite polynomials at the low scale $\mu_0 = 1$ GeV (black curve).
    The blue curve shows the analytic solution~\eqref{eq:h17mu} for $\mu = 2$ GeV, which is in good agreement with the numeric solution from discretisation of the momentum-space RG equation (blue dots).
    The latter is obtained using $N = 600$ points that are logarithmically distributed on the intervals $[-\Omega,-\eps]$ and $[\eps,\Omega]$, with $\eps = 10^{-9}$ GeV and $\Omega = 10^3$ GeV. 
    The strong coupling constant is evaluated with one-loop running and $n_f = 4$ quark flavors, using $\alpha_s(\mu_0) = 0.48$.
    We emphasize that the blue curve only includes the non-Abelian piece~\eqref{eq:gammanbar} of the anomalous dimension.}
    \label{fig:Hermite_evolution}
\end{figure}
We numerically test the derived solution of the RG equation using a simple model at the scale $\mu_0 = 1$ GeV. 
We follow~\cite{Benzke:2010js} and define
\begin{align}
\label{eq:h17definition}
   h_{17}(\omega_1,\mu_0) 
    = \int_{-\Delta}^{\bar{\Lambda}} d\omega \, g_{17}(\omega,\omega_1;\mu_0) \,,
\end{align}
with $\Delta = m_b - 2 E_0$ and $E_0$ is the energy cut in the photon spectrum.

As the evolution in the Abelian limit coincides with the well-known scale evolution of the leading shape function, we restrict ourselves to the non-Abelian piece in the following, such that at the scale $\mu > \mu_0$ one has\footnote{In general, including the Abelian part requires specifying a model for the shape function $g_{17}(\omega,\omega_1;\mu_0)$ instead of its integral $h_{17}(\omega_1;\mu_0)$.
For example, if one assumes at $\mu_0$ a factorisation of the form
\begin{equation*}
    g_{17}(\omega,\omega_1;\mu_0)=\hat{f}(\omega;\mu_0) \, h_{17}(\omega_1;\mu_0)\,,
\end{equation*}
with $\int_{-\Delta}^{\bar{\Lambda}} \! d\omega \, \hat{f}(\omega;\mu_0) = 1$, then the right-hand side of~\eqref{eq:h17mu} is simply multiplied by the factor 
\begin{equation*}
c(\mu,\mu_0)=\int_{-\Delta}^{\bar{\Lambda}}d\omega\int_\omega^{\bar{\Lambda}}\frac{d\omega^\prime}{\omega^\prime-\omega}U_{n}^{(17)}(\omega,\omega';\mu,\mu_0) \hat{f}(\omega;\mu_0)\,.
\end{equation*}
}
\begin{align}
\label{eq:h17mu}
    h_{17}(\omega_1;\mu)
    = \int_{-\Delta}^{\bar{\Lambda}} d\omega \, g_{17}(\omega,\omega_1;\mu)
    = \int_{-\infty}^\infty\frac{d\omega_1^\prime}{\vert\omega_1^\prime\vert}U_{\bar{n}}^{(17)}(\omega_1,\omega'_1;\mu,\mu_0) h_{17}(\omega_1;\mu_0)\,.
\end{align}
We emphasize again, that a serious numerical analysis -- including perturbative corrections to the jet and hard functions -- will be presented elsewhere.

A specific model for the function $h_{17}(\omega_1;\mu_0)$ has been introduced in~\cite{Gunawardana:2019gep}, and expands the function in Hermite polynomials multiplied by a Gaussian of width $\sigma$,
\begin{equation}
\label{eq:Hermitemodel}
    h_{17}(\omega_1;\mu_0) = \sum_n a_{2n} H_{2n} \left( \frac{\omega_1}{\sqrt{2} \sigma}\right) e^{-\frac{\omega_1^2}{2\sigma^2}} \,.
\end{equation}
Here the sum runs only over even integers to ensure that $h_{17}$ is an even function. 
In the following, we restrict ourselves to the first two terms $n \in (0,1)$, in which case the coefficients $a_0$ and $a_2$ can be related to positive moments of the function $h_{17}$~\cite{Gunawardana:2019gep},
\begin{equation}
  a_0 = \frac{\langle\omega_1^0 \, h_{17}\rangle}{\sqrt{2\pi} |\sigma|} \,, \qquad\qquad
  a_2 = \frac{\langle\omega_1^2 \, h_{17}\rangle - \sigma^2 \langle \omega_1^0 \, h_{17} \rangle}{4\sqrt{2\pi} |\sigma|^3} \,.
\end{equation}
As representative numerical values we use $\langle\omega_1^0 \, h_{17}\rangle = 0.25$ GeV$^2$, $\langle\omega_1^2 \, h_{17}\rangle = 0.1$ GeV$^4$, and $\sigma = 0.3$ GeV, which lie within the bounds quoted in~\cite{Gunawardana:2019gep}.
Fig.~\ref{fig:Hermite_evolution} shows this function at the two different scales $\mu_0 = 1$ GeV and $\mu = 2$ GeV, where the latter function is obtained analytically using the evolution function~\eqref{eq:CAevolutionfunctionsymm} for symmetric initial functions, and numerically by discretisation of the momentum-space RG equation (see caption of Fig.~\ref{fig:Hermite_evolution} for more details). 
The appearing integrals can be performed analytically, using e.g. \texttt{Mathematica}, and result in a sum of several higher Meijer-G functions, which we do not quote here.

%----------------------------------------------
\section{The exclusive counterpart \texorpdfstring{$\Phi_{\rm G}$}{PG}}
\label{sec:exclusive}
\begin{figure}[t]
    \centering
\includegraphics[width=0.37\textwidth]{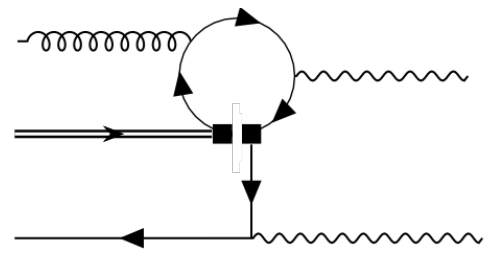}
    \caption{Leading order contribution from the operator $Q_1^c$ to the exclusive double radiative $\bar{B}_{d,s}\to\gamma\gamma$ decay.}
    \label{fig:exclusive_diagram}
\end{figure}
In this section, we analyse the soft function that has been introduced in~\cite{Qin:2022rlk} in the context of rare exclusive $\bar{B}_{s,d} \to \gamma \gamma$ decays, cf. Fig~\ref{fig:exclusive_diagram}, and whose renormalisation properties have recently been studied in~\cite{Huang:2023jdu}. 
It turns out that also in this case the analytic properties of the soft anomalous dimension allow one to considerably simplify scale evolution.
Provided the soft function is convoluted with appropriate jet functions, logarithms between the soft and the hard-collinear scale can be resumed by solving a ``reduced'' RG equation. 

The definition of the occurring soft function is rather similar to the one of $g_{17}(\omega,\omega_1;\mu)$ in~\eqref{eq:g17def}, and reads\footnote{We slightly changed the notation to be consistent with our definition of $g_{17}$.
In particular, we use the soft-momentum variables $(\omega,\omega_1)$ instead of $(\omega_1,\omega_2)$ used in~\cite{Huang:2023jdu}.}
\begin{align}
{\cal F}_B(\mu) \Phi_{\rm G}(\omega, \omega_1; \mu) = \frac{1}{2M_B} \int\frac{dr}{2\pi} \, e^{i\omega_1 r} \int\frac{dt}{2\pi}\,e^{i\omega t} \, \langle 0 | \mathcal{O}_G(t,r) | \bar B_v \rangle \,,
\label{eq:phi_G}
\end{align}
with the scale-dependent static decay constant ${\cal F}_B(\mu)$, and the HQET operator
\begin{equation}
\label{eq:exclusive_OG}
 \mathcal{O}_G(t,r) = (\bar q_{s} S_{n})(t n) \slashed{n} (S_{n}^{\dagger} S_{\bar n})(0) \, \gamma^{\perp}_{\alpha} \bar n_{\beta} \gamma_5 \,
(S_{\bar n}^{\dagger} \, g_s G_s^{\alpha\beta} S_{\bar n} )(r \bar n) \,
(S_{\bar n}^{\dagger} h_v)(0) \,.
\end{equation}

The most important difference to the shape function $g_{17}$ is that the soft function $\Phi_{\rm G}$ is defined at the amplitude level.   
Consequently, the soft quark-field $q_s$ is a light-quark field, $q = s,d$, and the operator contains a $\gamma_5$ Dirac matrix to ensure the correct parity. 
Furthermore, because all soft momenta are incoming, the signs of the exponentials are different from those in the definition of $g_{17}$ in~\eqref{eq:g17def}. 
As a consequence, the distributions $H_\pm$ transform according to 
\begin{align}
\label{eq:Hpmtransformations}
    &H_+(-\omega,-\omega^\prime)= H_-(\omega,\omega^\prime)+2\pi i\theta(\omega)\delta(\omega-\omega^\prime)\,, \nonumber \\
    &H_-(-\omega,-\omega^\prime)= H_+(\omega,\omega^\prime)-2\pi i\theta(-\omega)\delta(\omega-\omega^\prime) \,.
\end{align}

\subsection{Anomalous dimension}
\label{subsec:anomdimexclusive}
The anomalous dimension of the operator $\mathcal{O}_G$ was calculated in~\cite{Huang:2023jdu}, and reads
\begin{align}
\label{eq:exclusive_gamma} \Gamma_G(\omega,\omega_1,\omega',\omega'_1;\mu) = \, &\frac{\alpha_s C_F}{\pi}
\bigg[ \left( \ln{\frac{\mu}{\omega-i0}} -\frac{1}{2} \right) \delta(\omega - \omega^{\prime}) -  H_{+}(\omega, \omega^{\prime}) \bigg] \delta(\omega_1 - \omega_1^{\prime}) \nonumber \\[0.2em]
+ &\frac{\alpha_s C_A}{\pi} \bigg[ \left( \ln{\frac{\mu}{\omega_1-i0}} + \frac{i\pi}{2} \right) \delta(\omega_1 - \omega'_1) -  H_{+}(\omega_1, \omega'_1) \nonumber \\[0.2em]
&\phantom{\frac{\alpha_s C_A}{\pi}} \, +  \frac{\omega_1}{(\omega'_1)^2}
\left[ \theta(\omega_1) \theta(\omega'_1 - \omega_1) - \theta(-\omega_1) \theta(\omega_1 - \omega'_1) \right] \bigg] \delta(\omega - \omega') \nonumber \\[0.2em]
+ &\frac{\alpha_s}{\pi}\, \Delta\Gamma_G(\omega,\omega_1;\omega',\omega'_1;\mu) \,,
\end{align}
with
\begin{align}
\label{eq:exclusive_Deltagamma}
\Delta\Gamma_G = {\frac{i}{4}} {\frac{C_A}{\pi}}
\left[ \Delta H (\omega, \omega^{\prime}) 
- 2 i \pi \delta(\omega -  \omega^{\prime}) \right]
 \left[\Delta H(\omega_1, \omega_1^{\prime}) 
- 2 i \pi \delta(\omega_1 - \omega'_1)  \right] \,,
\end{align}
where we omitted the arguments of $\Delta\Gamma_G$ on the left-hand side for brevity, and defined $\Delta H \equiv H_+ - H_-$.
The Abelian contribution coincides with the well-known evolution kernel of the leading-twist $B$-meson light-cone distribution amplitude~\cite{Lange:2003ff}, or with its generalization to functions with support on the entire real axis~\cite{Beneke:2022msp}, respectively.
We note that the relation
\begin{equation}
    \ln \frac{\mu}{\omega - i0}\delta(\omega-\omega^\prime) - H_+(\omega,\omega') = \frac12 \ln \frac{\mu^2}{\omega^2}\delta(\omega-\omega^\prime) - \text{Re} \,H_+(\omega,\omega')
\end{equation}
implies that the Abelian part of the anomalous dimension is real-valued. 
The non-Abelian contributions can be extracted from the calculation of UV singularities presented in the previous section~\ref{subsec:CAterms} via the transformations~\eqref{eq:Hpmtransformations}, and by taking into account that the contribution~\eqref{eq:mixingtermdiagb} from diagram $(b)$ is UV-finite in the light-quark case.

The RG equation with the anomalous dimension~\eqref{eq:exclusive_gamma} has been solved in~\cite{Huang:2023jdu}.
Notably, the distribution $H_-$ generates a non-vanishing support for negative values of the soft momentum variables~\cite{Beneke:2022msp}; a somewhat unfamiliar feature for amplitude-level soft functions.
Using similar arguments as in the previous section, we will now argue that the analytic properties of the soft anomalous dimension and the jet functions allow one to solve a ``reduced'' RG equation that features two important simplifications:
\begin{itemize}
    \item First, as factorisation in SCET would indicate, the dependence on the soft momenta $(\omega,\omega_1)$ associated with the two light-cone directions $n^\mu$ and $\bar{n}^\mu$ decouples, 
    \item and second, a positively-supported function at the low scale $\mu_0$ does \emph{not} acquire support for negative values at $\mu > \mu_0$ in either of the two variables. 
\end{itemize}
The latter property has the important consequence that no additional complex phases are generated through the convolution integrals in inverse moments.
Furthermore, we then find that the asymptotic behavior for small arguments is consistent with the usual conformal spin of the soft fields.
Using a specific example, these features are discussed in greater detail below.

The calculation of the anomalous dimension for an amplitude-level soft function proceeds via standard time-ordered Feynman diagrams that do not contain any cut propagators.
However, given that~\eqref{eq:Hpmidentity} is equivalent to
\begin{equation}
    \frac{1}{\omega - \omega' - i0} = H_+(\omega,\omega') - H_-(\omega,\omega') \,,
\end{equation}
and the jet functions for the exclusive $\bar{B}_{d,s} \to \gamma \gamma$ process have singularities or branch-cuts located in the upper half-plane (i.e. depend on $\omega - i0$ and $\omega_1 - i0$, respectively), implies that we can identify the two distributions $H_+(\omega,\omega') = H_-(\omega,\omega')$.\footnote{The $i0$ prescriptions of the jet functions differ in inclusive and exclusive processes due to the different kinematics.}
For the special case of jet functions of the form $\sim (\omega - i0)^\eta$ this has already been realised in~\cite{Beneke:2022msp}.
Here we give an explanation for this observation, and generalize this statement to jet functions with similar analytic properties.

To give a more formal reasoning, it is important to note that distribution-valued objects are always defined on a function space with certain properties that are required e.g. for the convergence of integrals.
This function space is typically defined in the primed variables  $(\omega',\omega'_1)$, as these are the integration variables in the RG equation.
However, as was also discussed in~\cite{Beneke:2022msp} in the context of these specific distributions, the soft function itself can be considered as a distribution-valued object in the variables $(\omega,\omega_1)$, see for example the fixed-order expressions for the $B_c$-meson light-cone distributions amplitude in the non-relativistic approximation in~\cite{Bell:2008er}. 
In this sense, the identity $H_+ = H_-$ holds on a function space in $(\omega,\omega_1)$ with two properties: First, all singularities and branch cuts of the test functions are located in the upper half-plane, and second, the functions are such that the residue theorem can be applied.
The latter means that they should
fall of fast enough for large $(\omega,\omega_1)$, and the integration domain should extend over the entire real axis.     
From dimensional arguments, and due to the directed flow of large energy, both of these properties are fulfilled by the relevant jet functions for the $\bar{B}_{d,s} \to \gamma \gamma$ process.
For example, the leading-order jet function in the $n^\mu$ direction is simply given by the inverse hard-collinear propagator $\sim (\omega - i0)^{-1}$, and similarly the jet function in the $\bar{n}^\mu$ direction is given by $(\omega_1 - i0)^{-1}$ at leading order and in the massless case $m_u = 0$. 

We emphasize again that we have checked the cancellation of all singularities in $1/\eps$ between the different elements of the factorisation theorem at next-to-leading order and in the limit $m_u = 0$.
The relevant two-loop $\bar{n}$-jet function is the same as for the inclusive $\bar{B} \to X_s \gamma$ decay up to the different sign of the $i0$ prescription. 

To conclude this discussion let us stress that the anomalous dimension of a given operator is well-defined and unambiguous.
Identifying $H_+ = H_-$ is allowed whenever the soft operator is convoluted with the jet functions, and can hence simplify the resummation of logarithms between the soft and the hard-collinear scale. 
As soon as one is concerned about the soft function itself, however, it should be considered as a complex function with certain analytic properties, and one is not allowed to apply any simplifying identities.

Coming back to the anomalous dimension~\eqref{eq:exclusive_gamma}, we can now replace
\begin{equation}
 \frac{\alpha_s}{\pi} \Delta\Gamma_G \to -\frac{\alpha_s C_A}{\pi} i \pi \delta(\omega-\omega^\prime)\delta(\omega_1-\omega_1^\prime) \,,
\end{equation}
which allows us to write
\begin{equation}
\label{eq:exclusive_anomdim_factorisation}
\Gamma_{G}(\omega,\omega_1,\omega',\omega'_1;\mu)
 = \frac{\alpha_s}{\pi} \Big\{ C_F \, \delta(\omega_1 - \omega'_1) \Gamma_{n}(\omega,\omega';\mu)
+ C_A \, \delta(\omega - \omega') \Gamma_{\bar{n}}(\omega_1,\omega'_1;\mu) \Big\} \,,  
\end{equation}
as one would naively expect from factorisation in SCET. 
The individual $\mathcal{O}(\alpha_s)$ anomalous dimensions in the two collinear sectors read
\begin{align}
\label{eq:gamma_nnbar_exclusive}
    \Gamma_{n}(\omega,\omega';\mu) = \, &\frac{1}{2}\left(\ln \frac{\mu^2}{\omega^2} - 1\right) \delta(\omega-\omega')
    - \text{Re} \,H_+(\omega,\omega^\prime) \,, \\[0.2em]
     \Gamma_{\bar{n}}(\omega_1,\omega'_1;\mu) = \, &\frac{1}{2}\left(\ln \frac{\mu^2}{\omega_1^2}-i\pi\right) \, \delta(\omega_1-\omega'_1) -  \text{Re}\,H_+(\omega_1,\omega'_1)\nonumber\\[0.2em]
    &+ \frac{\omega_1}{(\omega'_1)^2} \big[ \theta(\omega_1)\theta(\omega'_1 - \omega_1) - \theta(-\omega_1) \theta(\omega_1 - \omega'_1) \big] \,,
\end{align}
and again one can solve two separate RG equations related to the two variables $\omega$ and $\omega_1$, respectively.

\subsection{Solution to the ``reduced'' RG equation}
The derivation of the solutions to these ``reduced'' RG equations for the Abelian and the non-Abelian part is very similar.
In the following we present some details on the latter case, for which we follow the same procedure as in the previous section, but we quote the complete result for the evolution functions below. 
Performing the usual split into a positive-support and negative-support branch yields the Mellin-space equations
\begin{align}
    \left( \frac{d}{d\ln \mu} -\eta_{1} \right) \widetilde{\Phi}_{G}^>(\omega,\eta_{1};\mu) = \,&\frac{\alpha_s C_{A}}{\pi} \Big[ -H_{1-\eta_{1}}-H_{\eta_{1}} -\partial_{\eta_{1}} + \frac{i\pi}{2}\Big]\widetilde{\Phi}_{G}^>(\omega,\eta_{1};\mu)\nonumber\\
    &+ \frac{\alpha_s C_{A}}{\pi} \Gamma(-\eta_{1}) \Gamma(1+\eta_{1})  \widetilde{\Phi}_{G}^<(\omega,\eta_{1};\mu) \,, \nonumber \\[0.2em]
    \left( \frac{d}{d\ln \mu} -\eta_{1} \right) \widetilde{\Phi}_{G}^<(\omega,\eta_{1};\mu) = \, &\frac{\alpha_s C_{A} }{\pi} \Big[ - H_{-1-\eta_{1}}-H_{1-\eta_{1}} -\partial_{\eta_{1}} + \frac{i\pi}{2} \Big]\widetilde{\Phi}_{G}^<(\omega,\eta_{1};\mu) \,,
    \label{eq:exclusivenonAbelianMellinspaceRGEs}
\end{align}
which again hold for $-1<\text{Re}(\eta_1)<0$.
Notably, the equation for the negative-support branch is diagonal, because the anomalous dimension $\Gamma_{\bar{n}}(\omega_1,\omega'_1;\mu)$ in~\eqref{eq:gamma_nnbar_exclusive} does not contain a respective mixing piece, which would arise for example from an $H_-$ distribution. 
This implies that a function with no negative support at the scale $\mu_0$ does \emph{not} develop support for negative $\omega_1$ through scale evolution, which can already be seen from~\eqref{eq:gamma_nnbar_exclusive} by noting that the distribution $H_+$ reduces to the standard Lange-Neubert kernel~\cite{Lange:2003ff},
\begin{equation}  
H_+(\omega,\omega') \to F(\omega,\omega^\prime)=\omega\left[\frac{\theta(\omega^\prime-\omega)}{\omega^\prime(\omega^\prime-\omega)}\right]_++\left[\frac{\theta(\omega-\omega^\prime)}{\omega-\omega^\prime}\right]_+\,, \qquad \text{for } \omega,\omega'>0 \,.
\end{equation}

The solution for $\widetilde{\Phi}_{G}^<(\omega,\eta_{1};\mu)$ can be obtained straightforwardly and reads
\begin{align} 
    \widetilde{\Phi}_{G}^<(\omega,\eta_{1};\mu) &= e^{V_G+2\gamma_E a_1} \left( \frac{\mu}{\mu_0} \right)^{\!\!\eta_{1}} \frac{\Gamma(-\eta_{1})\Gamma(2-\eta_{1})}{\Gamma(-\eta_{1}-a_1)\Gamma(2-\eta_{1}-a_1)} \,\widetilde{\Phi}_{G}^<(\omega,\eta_{1}+a_1;\mu_0) \; , 
\end{align}
where $a_1$ is defined in~(\ref{eq:VaQCDCA}) and the evolution factor
\begin{align}
    V_G(\mu,\mu_0)  =&-\int_{\mu_0}^\mu \frac{d\mu'}{\mu'} \frac{\alpha_s(\mu') C_{A} }{\pi} \bigg[  \ln \frac{\mu'}{\mu_0} -\frac{i\pi}{2}  \bigg] \; ,  
\end{align}
differs from the definition of $V_1$ in~(\ref{eq:VaQCDCA}) due to the presence of the imaginary part $-i\pi/2$ in the square brackets.
The system~\eqref{eq:exclusivenonAbelianMellinspaceRGEs} can be diagonalized using the following linear combination as the second independent function,
\begin{equation}
\widetilde{\Phi}_{G}(\omega, \eta_{1};\mu)\equiv \int_{-\infty}^{+\infty} \! \frac{d\omega_{1}}{\omega_{1}-i0} \left( \frac{\mu}{\omega_{1}-i0} \right)^{\eta_{1}}\Phi_G(\omega, \omega_{1};\mu) =\widetilde{\Phi}_{G}^>(\omega,\eta_{1};\mu)-\widetilde{\Phi}_{G}^<(\omega,\eta_{1};\mu)e^{i\pi\eta_{1}},
\end{equation}
which is solved by
\begin{align} 
    \widetilde{\Phi}_{G}(\omega, \eta_1;\mu) &= e^{V_G+2\gamma_E a_1} \left( \frac{\mu}{\mu_0} \right)^{\!\!\eta_1} \frac{\Gamma(2-\eta_1)\Gamma(1+\eta_1+a_1)}{\Gamma(1+\eta_1)\Gamma(2-\eta_1-a_1)} \,\widetilde{\Phi}_{G}(\omega, \eta_1+a_1;\mu_0) \,. 
\end{align}
Performing the inverse Mellin transform, the momentum-space evolution functions can again be expressed in terms of Meijer-G functions. 
Including also the Abelian piece, the solution to the RG-equation for $\Phi_G$ -- restricted to the function space of relevant jet functions -- then reads
\begin{equation}
\Phi_G(\omega,\omega_1;\mu) = \int \frac{d\omega'}{\omega'} \int \frac{d\omega'_1}{\omega'_1} \, 
U_n^{(G)}(\omega,\omega';\mu,\mu_0) 
U_{\bar{n}}^{(G)}(\omega_1,\omega'_1;\mu,\mu_0) \, \Phi_G(\omega',\omega_1^\prime;\mu_0) \,. \label{eq:final_evolution_exclusive}
\end{equation}
The evolution functions are
\begin{align}
 U_n^{(G)}(\omega,\omega';\mu,\mu_0) = -e^{V+2\gamma_E a} \bigg(\frac{\mu_0}{\omega'-i0}\bigg)^{a}\bigg[&\theta(-\omega')e^{-i\pi a}
 G^{2,0}_{2,2} \bigg( { {-a \, ,1-a}\atop {1 \,\,, \,\,\,\,\,\,0 }} \bigg| \frac{\omega}{\omega'} \bigg)\nonumber \\[0.2em]
 - &\theta(\omega)G^{1,1}_{2,2} \bigg( { {-a \, ,1-a}\atop {1 \,\,, \,\,\,\,\,\,0 }}   \bigg|\frac{\omega}{\omega'}\bigg) \bigg] \,,
\end{align}
with $V = V(\mu,\mu_0)$ and $a = a(\mu,\mu_0)$ defined in~\eqref{eq:VaQCDCF}, and
\begin{align}
 U_{\bar{n}}^{(G)}(\omega_1,\omega'_1;\mu,\mu_0) = -e^{V_G+2\gamma_E a_1} \left(\frac{\mu_0}{\omega_{1}^\prime-i0}\right)^{a_1}\bigg[&\theta(-\omega_1^\prime) e^{-i\pi a_1}G^{2,0}_{2,2} \bigg( { {-a_1 \, ,2-a_1}\atop {2 \,\,, \,\,\,\,\,\,0 }}   \bigg| \frac{\omega_1}{\omega'_1} \bigg)
 \nonumber \\[0.2em]
 -&\theta(\omega_1) G^{1,1}_{2,2} \bigg( { {-a_1 \, ,2-a_1}\atop {2 \,\,, \,\,\,\,\,\,0 }}   \bigg| \frac{\omega_1}{\omega'_1} \bigg) \bigg] \,.
\end{align}
Again, the Meijer-G functions can be reduced to hypergeometric functions in certain integration domains. 
For $\omega/\omega' > -1$ the identity
\begin{align}
\label{eq:G2211identity}
    G^{1,1}_{2,2} \bigg( { {-x \, ,\alpha-x}\atop {\alpha \,\,, \,\,\,\,\,\,0 }}   \bigg| \frac{\omega}{\omega'} \bigg)
    = \, &\frac{\Gamma(1+x+\alpha)}{\Gamma(-x)\Gamma(1+\alpha)} \left( \frac{\text{min}(\omega,\omega')}{\text{max}(\omega,\omega')} \right)^\alpha \left( \frac{\omega'}{\text{max}(\omega,\omega')} \right)^{x+1-\alpha} \nonumber \\[0.2em]
    &\times {}_2F_1\left(x+1,x+1+\alpha;\alpha+1;\frac{\text{min}(\omega,\omega')}{\text{max}(\omega,\omega')}\right)
\end{align}
holds, where an implicit $+i0$ prescription is understood in the brackets of the first line. The second appearing Meier-G function vanishes for $|\omega/\omega'|\, > 1$,
\begin{align}
    G^{2,0}_{2,2} \bigg( { {-x \, ,\alpha-x}\atop {\alpha \,\,, \,\,\,\,\,\,0 }}   \bigg| \frac{\omega}{\omega'} \bigg)
    = 0 \,.
\end{align}

\subsection{Phenomenological implications}
We now consider the special case where $\Phi_{\rm G}(\omega, \omega_1, \mu_0) \sim \theta(\omega) \theta(\omega_1)$ at the low scale $\mu_0$.
The solution to the ``reduced'' RG-equation then simplifies to
\begin{align} 
\Phi_G(\omega,\omega_1;\mu) = \, &\theta(\omega) e^{V+2\gamma_E a} \int_0^\infty \frac{d\omega'}{\omega'} \, \bigg(\frac{\mu_0}{\omega'}\bigg)^{a} G^{1,1}_{2,2} \bigg( { {-a \, ,1-a}\atop {1 \,\,, \,\,\,\,\,\,0 }}   \bigg|\frac{\omega}{\omega'}\bigg) \\[0.2em]
\times \, &\theta(\omega_1) e^{V_G +2\gamma_E a_1}  \int_0^\infty \frac{d\omega'_1}{\omega'_1} \,  \left(\frac{\mu_0}{\omega_{1}^\prime}\right)^{a_1}
G^{1,1}_{2,2} \bigg( { {-a_1 \, ,2-a_1}\atop {2 \,\,, \,\,\,\,\,\,0 }}   \bigg| \frac{\omega_1}{\omega'_1} \bigg) \, \Phi_G(\omega',\omega_1^\prime;\mu_0) \,, \nonumber
\end{align}
and, since all variables are positive, we can substitute the Meijer-G functions using the identity~\eqref{eq:G2211identity}.

The solution in the Abelian limit then corresponds to the well-known solution of the Lange-Neubert kernel for the leading-twist $B$-meson distribution amplitude~\cite{Bell:2013tfa}, and takes a similar form for the non-Abelian piece, but with the parameter $\alpha = 2$ in~\eqref{eq:G2211identity}.
This in particular implies that $\Phi_G(\omega,\omega_1;\mu)$ falls off linearly in $\omega$, but quadratically $\omega_1$ for small values,
\begin{equation}
 \Phi_G(\omega,\omega_1;\mu) \sim \omega \omega_1^2 \,, \qquad \text{for small } \omega, \omega_1 \,.
\end{equation}
For large arguments, on the other hand, the function behaves as $\sim \omega_1^{-a_1-\text{min}(1,\xi_1)}$ for $\omega_1 \to \infty$, and as $\sim \omega^{-a-\text{min}(1,\xi)}$ for $\omega \to \infty$, if the initial function falls off power-like with exponents $\xi_1$ and $\xi$.
Hence, convolution integrals in factorisation theorems again converge
as long as the $\mu$ is not chosen to be unphysically large.
Interestingly, due to the factorisation of the two light-cone directions in~\eqref{eq:exclusive_anomdim_factorisation}, the asymptotic behavior for small momenta is consistent with the conformal spins of the light-quark and gluon fields~\cite{Braun:1989iv} (see also~\cite{Braun:2017liq}) that appear in the operator~\eqref{eq:exclusive_OG}.
Although this is merely an observation at this point, a connection between the conformal spin of the fields and the asymptotic behavior for such multi-light-cone soft functions would be important for studying the convergence of convolution integrals.
As such soft functions will become relevant in a variety of inclusive and exclusive processes at next-to-leading power, a more rigorous understanding of this feature would be important. 

As a concrete example, we study the same simple exponential model that was used in~\cite{Huang:2023jdu},
\begin{equation}
\label{eq:exp_model_exclusive}
    \Phi_G^{\text{exp}}(\omega,\omega_1,\mu_0) = \frac{\lambda_E^2+\lambda_H^2}{6}\frac{\omega\omega_1^2}{\omega_0^5}e^{-\frac{\omega+\omega_1}{\omega_0}}\theta(\omega)\theta(\omega_1) \,,
\end{equation}
where the hadronic quantities $\lambda_E$ and $\lambda_H$ parametrize hadronic matrix elements of the chromo-electric and chromo-magnetic
fields, respectively~\cite{Grozin:1996pq}.
In particular, as the dependence on $\omega$ and $\omega_1$ factorizes, we can also express the function at the scale $\mu > \mu_0$ as a product of functions that only depend on either one of the two variables.
The integrals can easily be performed analytically, and result in
\begin{align}
\label{eq:exp_model_exclusive_solution}
    \Phi_G^{\text{exp}}(\omega,\omega_1,\mu) = \, &\theta(\omega) \frac{\lambda_E^2+\lambda_H^2}{6\omega_0^2}\frac{\omega}{\omega_0}e^{V+2\gamma_E a}  \bigg(\frac{\mu_0}{\omega_0}\bigg)^{a}\Gamma\left(2+a\right) \, {}_1F_1\left(2+a;2;-\frac{\omega}{\omega_0}\right) \nonumber\\[0.2em]
    \times \, &\theta(\omega_1) \frac{\omega_1^2}{2\omega_0^2}e^{V_G+2\gamma_E a_1}\bigg(\frac{\mu_0}{\omega_0}\bigg)^{a_1}\Gamma(3+a_1)\, {}_1F_1\left(3+a_1;3;-\frac{\omega_1}{\omega_0}\right) \,.
\end{align}

Importantly, inverse moments of the form
\begin{equation}
    \int_{-\infty}^{+\infty} \frac{d\omega}{\omega - i0} \, \int_{-\infty}^{+\infty} \frac{d\omega_1}{\omega_1 - i0} \, \Phi_G(\omega,\omega_1;\mu) 
\end{equation}
are no longer integrated over the poles at $\omega = 0$ or $\omega_1 =0$, respectively. 
Thus, the convolution integrals do no longer give rise to complex phases.  
As a cross check of the ``reduced'' RG evolution, for which the integrations can be performed analytically, we have explicitly verified that the expression for the evolved function in~\eqref{eq:exp_model_exclusive_solution} leads to the same double inverse moment as the result provided in~\cite{Huang:2023jdu} using the full anomalous dimension,
\begin{align}
    &\int_{0}^{\infty}\frac{d\omega}{\omega}  \, \int_{0}^{\infty}\frac{d\omega_1}{\omega_1} \, \Phi_G(\omega,\omega_1;\mu) 
    = \int_{-\infty}^{+\infty}\frac{d\omega}{\omega-i0} \, \int_{-\infty}^{+\infty}\frac{d\omega_1}{\omega_1-i0} \, \Phi_G^{\text{\cite{Huang:2023jdu}}}(\omega,\omega_1;\mu) \nonumber \\[0.2em]
    = \, &\frac{\lambda_E^2+\lambda_H^2}{6\omega_0^2} e^{V_G + V + 2\gamma_E (a_1 + a)} \left( \frac{\mu_0}{\omega_0} \right)^{a_1 + a} \Gamma(1 + a_1) \Gamma(1 + a) \,.
\end{align}

\begin{figure}[t]
    \centering
    \includegraphics[width=0.75\textwidth]{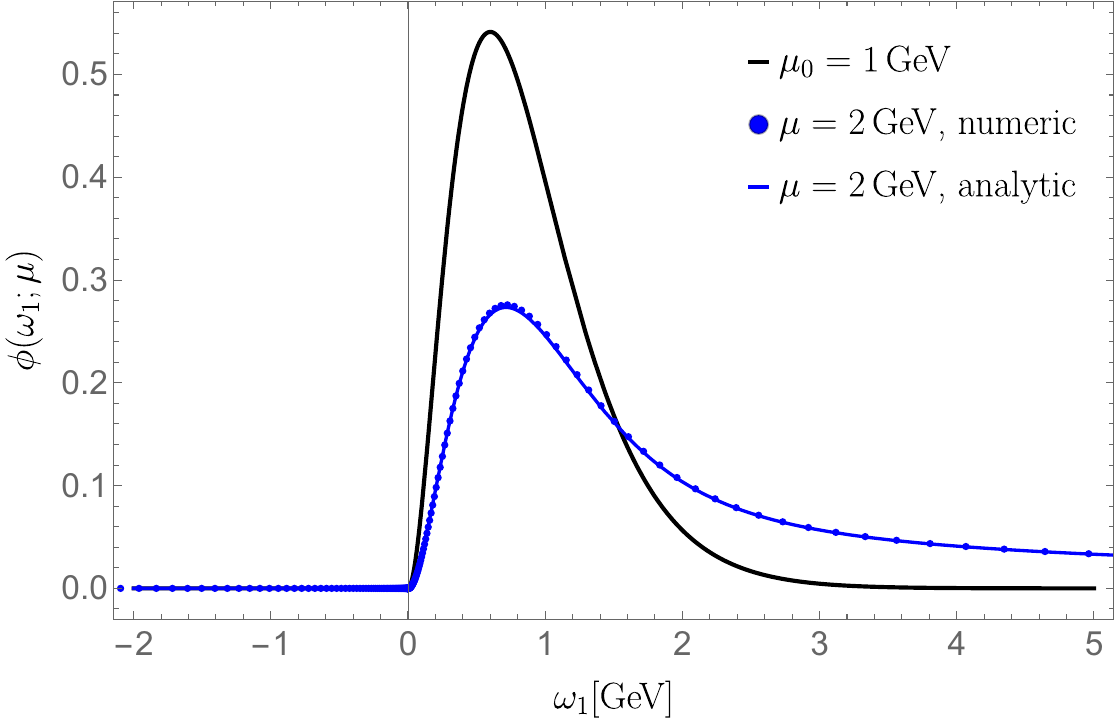}
    \caption{Scale evolution of the function $\phi(\omega_1;\mu)$, using the exponential model~\eqref{eq:phiexpmodel} with $\omega_0 = 0.3$ GeV at the low scale $\mu_0 = 1$ GeV (black curve).
    The blue curve shows the analytic solution for $\mu = 2$ GeV, which again is in good agreement with the numeric solution from discretisation of the momentum-space RG equation (blue dots). 
    Details on the discretisation and the strong coupling constant are explained in the caption of Fig.~\ref{fig:Hermite_evolution}.}
    \label{fig:phiG_evolution}
\end{figure}

Lastly, we again verify our analytic solution numerically through discretisation of the momentum-space RG equation in Fig.~\ref{fig:phiG_evolution}. To do so, we define a function $\phi(\omega_1;\mu)$ that fulfills a RG equation with anomalous dimension
\begin{align}
    \frac{\alpha_s C_A}{\pi} \! \left(\frac{1}{2} \ln \frac{\mu^2}{\omega_1^2} \, \delta(\omega_1-\omega'_1) -  \text{Re}\,H_+(\omega_1,\omega'_1)
    + \frac{\omega_1}{(\omega'_1)^2} \big[ \theta(\omega_1)\theta(\omega'_1 - \omega_1) \! - \! \theta(-\omega_1) \theta(\omega_1 - \omega'_1) \big]\right),
\end{align}
that corresponds to $\Gamma_{\bar{n}}(\omega_1,\omega'_1;\mu)$ up to a local complex phase which can be factored out. 
We choose
\begin{equation}
\label{eq:phiexpmodel}
    \phi(\omega_1,\mu_0) = \theta(\omega_1) \frac{\omega_1^2}{\omega_0^2}e^{-\frac{\omega_1}{\omega_0}}
\end{equation}
as an initial condition, such that its analytic expression at the scale $\mu$ precisely results in the second line of~\eqref{eq:exp_model_exclusive_solution}, but with $V_G$ replaced by its real part.

%----------------------------------------------
\section{Conclusions}
\label{sec:conclusion}
In this paper, we presented the initial steps towards incorporating $\mathcal{O}(\alpha_s)$ radiative corrections to the resolved-photon contribution from the $Q^c_1 - Q_{7\gamma}$ interference in $\bar{B}\to X_s\gamma$ and $\bar{B}\to X_s\ell^+\ell^-$ decays.
These contributions are expected to reduce the sizeable scale ambiguity and charm-mass dependence of the leading-order result, which currently provides a significant source of uncertainties in these decays. 
More precisely, here we studied the renormalisation of the subleading shape function $g_{17}(\omega,\omega_1;\mu)$, which describes the non-perturbative soft physics in the factorisation of these contributions.
As next-to-leading (NLO) order corrections to the quark jet function and the hard matching coefficients are well-known, the last missing piece for a complete NLO analysis consists of the calculation of the $\mathcal{O}(\alpha_s)$ corrections to the jet function that describes the conversion of two anti-hard-collinear quarks to a soft gluon and an energetic photon. 
We have computed the singularities in $1/\eps$ of these two-loop corrections in the limit of vanishing quark mass, and they serve as an important cross check for the consistency of the factorisation formula at NLO, but we relegate a detailed analysis to a forthcoming publication. 

On a technical level, the underlying operator in Heavy-Quark Effective Theory is composed of two heavy-quark fields with light-like separation in the $n^\mu$ direction, as well as a soft gluon field that is smeared along the opposite $\bar{n}^\mu$ light-cone.
Computing the ultraviolet singularities of matrix elements of this operator with standard time-ordered products results in terms in the anomalous dimension that superficially seem to be in conflict with soft-collinear factorisation, as the two collinear sectors are not decoupled. 
However, these contributions vanish upon taking the necessary restricted cuts for the $\bar{B} \to X_s \gamma$ decay rate at the level of the soft function.
Hence, the shape function $g_{17}$ should be defined by an operator that distinguishes the fields belonging to the amplitude and the complex-conjugate amplitude, which can be done at the level of the path integral using the Keldysh formalism. 

The resulting anomalous dimension can be expressed as a sum of two terms, each of which only depends on the soft momentum variable associated with one of the two light-cone directions, $\omega$ or $\omega_1$, respectively. 
As a consequence, the renormalisation group (RG) equation decomposes into two separate equations associated with the two collinear sectors. 
We present analytic solutions to these equations in momentum-space, and qualitatively study some phenomenological implications of the scale evolution of $g_{17}(\omega,\omega_1;\mu)$. 

Motivated by the observations made during this analysis, we further studied a related amplitude-level soft function $\Phi_G(\omega,\omega_1;\mu)$, that describes the non-perturbative long-distance dynamics in penguin contributions to exclusive $\bar{B}_{d,s} \to \gamma \gamma$ decays. 
The renormalisation of this soft function has recently been studied in~\cite{Huang:2023jdu}.
We confirm the reported result for the anomalous dimension, which contains
products of non-trivial distributions in the variables $\omega$ and $\omega_1$. 
Again, such terms seem to spoil soft-collinear factorisation, but contrary to the inclusive modes they cannot be eliminated by taking cuts.
Still, we found that they are irrelevant in factorisation theorems, i.e. after convolution with the jet functions, due the analytic properties. 
More precisely, the poles and branch cuts in the complex $\omega$-plane (or $\omega_1$-plane) of the soft anomalous dimension and the jet functions are on the same side, such that the integration contour can be closed to avoid these singularities.

Such multi-light-cone soft functions should hence be considered as complex-valued functions with certain analytic properties that can be exploited and might lead to simplifications in factorisation theorems. 
We have here shown that this can be done at the level of the one-loop anomalous dimension.

Based on this insight, we solve a much simpler ``reduced'' RG equation for the soft function $\Phi_G(\omega,\omega_1;\mu)$, which again decomposes into two separate equations in the variables $\omega$ and $\omega_1$. 
Interestingly, we then find that scale evolution of this function is almost ``Lange-Neubert like'' in both variables.
This considerably simplifies the resummation of large logarithms in RG-improved perturbation theory.
An important consequence is that the negative support, induced by some of the distributions in the full anomalous dimension, can be avoided, provided a model with only positive support is used at the low scale $\mu_0$.\footnote{However, we should emphasize that at this point it remains an open question whether or not non-perturbative effects at the hadronic scale $\mu_0 \approx 1$ GeV create a support for negative values of $\omega$ or $\omega_1$.} 
Furthermore, we also find that in this case the asymptotic behavior for small arguments is consistent with the conformal spins of the light-quark and gluon fields, which -- if true in general -- would be important to study the convergence of convolution integrals in many processes at subleading power.

\section*{Acknowledgements}
We thank Thomas Becher, Martin Beneke, Michael Benzke and Patrick Hager for discussions.
P.B. is grateful to Vladyslav Shtabovenko for providing invaluable guidance on using \texttt{FeynCalc}.
All three authors have been supported by  the  Cluster  of  Excellence  ``Precision  Physics,  Fundamental
Interactions, and Structure of Matter" (PRISMA$^+$ EXC 2118/1) funded by the German Research Foundation (DFG) within the German Excellence Strategy (Project ID 390831469). 
The research of P.B. was funded by the European Union's Horizon 2020 research and innovation programme under the Marie Sk\l{}odowska-Curie grant
agreement No.101146976, and the European Research Council (ERC) under the European Union’s Horizon 2022 Research and Innovation Program (ERC Advanced Grant agreement No.101097780, EFT4jets).
P.B. thanks the Munich Institute for Astro-, Particle and BioPhysics (MIAPbP) for hospitality during
the final stages of this work.
T.H. thanks the CERN theory group for its hospitality during his regular visits to CERN where part of the work was done.

\begin{appendix}

\section{On the support of matrix elements of \texorpdfstring{$\mathcal{O}_{17}$}{O17}}
\label{app:support}
\begin{figure}[t]
    \centering
    \includegraphics[width=0.52\textwidth]{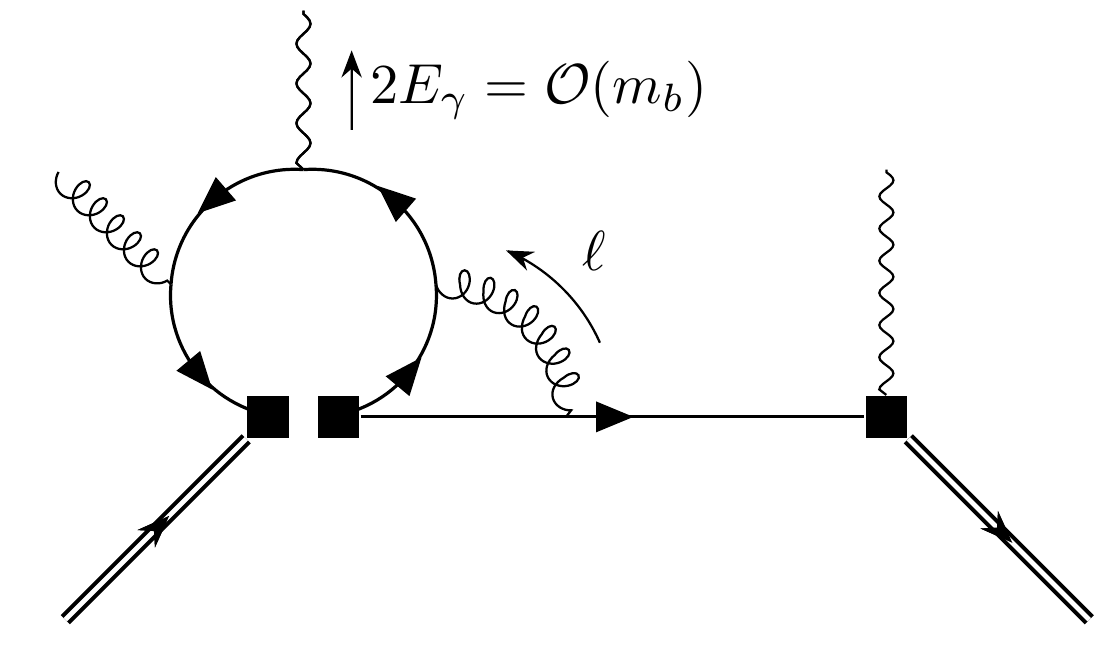}
    \caption{Example full-theory (i.e.~QCD) diagram that contributes to~\eqref{eq:ZfactorCAdiagramgnnbarWL} if the virtual gluon is soft.
    The outgoing photon has large energy of $\mathcal{O}(m_b)$ in the $\bar{n}^\mu$ direction, which is set to infinity in the heavy-quark limit. 
    Large-energy flows ``upwards'' from the four-fermion operator $Q_1^{(q)}$ to the outgoing photon, which implies that the $\ell_+$-component of the soft-gluon momentum has no upper bound for $E_\gamma \to \infty$.  
    The $\delta$-function from the Feynman rule then enforces $\omega = \omega'+\ell_+$ and creates a support for $\omega \to \infty$.}
    \label{fig:full_theory_diagram}
\end{figure}

In this appendix we argue that matrix elements of the operator $\mathcal{O}_{17}$ must be integrated over the entire real axis in both light-cone momentum projections. 
In particular, the variable $\omega$ associated with the collinear $n^\mu$ direction must be integrated up to $+\infty$ for a consistent renormalisation in the $\overline{\text{MS}}$-scheme.
This can be seen, for example, by the following argument, which has already been presented in~\cite{Beneke:2022msp} in a similar form.
Upon taking the local limit $t \to 0$, the finite-length Wilson line in the $n^\mu$ direction collapses to unity.
In momentum space, taking $t \to 0$ implies integration over $\omega$.
The left-hand side of~\eqref{eq:ZfactorCAdiagramgnnbarWL} then obviously vanishes, as it should, because the difference of $\delta$-functions represents a finite-length Wilson line.
This, however, only happens if $\omega$ is integrated up to $+\infty$.
An easy way to see this is by integrating the left-hand side of~\eqref{eq:ZfactorCAdiagramgnnbarWL} first in $\omega_1$ with the weight function $(\omega_1 + i0)^{-1}$, which -- neglecting irrelevant prefactors that do not depend on $\omega$ or $\omega'$ -- leads to
\begin{align}
    \propto \Gamma(\eps) \left( \frac{\theta(\omega - \omega')(\omega - \omega')^{-\eps}}{\omega - \omega' + \Delta} - \delta(\omega - \omega') \Gamma(\eps) \Gamma(1-\eps) \Delta^{-\eps} \right) \,.
\end{align}
This expression is in fact almost identical to the one discussed in (2.6) of~\cite{Beneke:2022msp}, but with the roles of $\omega$ and $\omega'$ interchanged, and it is easy to check that it does not vanish under a cut-off integral, and would result in additional divergences. 

The support $\omega \in (-\infty,+\infty)$ is hence required for consistency with the local limit $t\to0$, but it can also be explained by physics arguments.
Consider, for example, the diagram in Fig.~\ref{fig:full_theory_diagram} in the full theory (i.e. QCD) in Feynman gauge.
In the momentum region where the exchanged gluon with momentum $\ell$ becomes soft, the couplings to the two (anti-)hard-collinear quark lines lead to soft Wilson lines, which in the effective theory arise from the decoupling transformation of the respective quark fields.
The shown diagram is thus part of the HQET diagram $(g)$ in Fig.~\ref{fig:g17_diagrams}, and contributes to~\eqref{eq:ZfactorCAdiagramgnnbarWL}.
Recall that the kinematics of the $b \to s \gamma$ process is such that the outgoing $s$-quark field is hard-collinear, i.e. has large energy of $\mathcal{O}(m_b)$ in the $n^\mu$ direction, whereas the outgoing photon has a large energy $E_\gamma \sim \mathcal{O}(m_b)$ in the opposite $\bar{n}^\mu$ direction.
The large-energy flow along the $\bar{n}^\mu$ light-cone in Fig.~\ref{fig:full_theory_diagram} is thus from the heavy $b$-quark, upwards through the quark loop to the outgoing photon.
In the effective theory, however, the large energy scale $m_b$ is integrated out and set to infinity.
The light-cone component $\ell_+$ of the quark-momentum can hence also be of $\mathcal{O}(E_\gamma \sim m_b \to \infty)$, and the first $\delta$-function in the second line of~\eqref{eq:ZfactorCAdiagramgnnbarWL} enforces $\omega = \omega'+\ell_+$. 
This means that, for any $\omega'$ that is chosen as an initial value, radiative corrections will generate a support in $\omega$ up to $+\infty$.

\end{appendix}

\bibliographystyle{JHEP}
\bibliography{bibliography}

\end{document}